\documentclass[aps,amsmath,amssymb,reprint,groupedaddress]{revtex4-1}
\usepackage{indentfirst} %
\usepackage{amsmath,amssymb}   
\usepackage{latexsym} 
\usepackage{graphicx} %
\usepackage{subfigure}   
\usepackage{color}
\usepackage{CJK}
\usepackage{leftidx}
\usepackage{bm}

\usepackage[left=25mm,right=25mm,bottom=30mm,top=25mm]{geometry}

\begin{document}

\title{Electro-Neutral Models for dynamic Poisson-Nernst-Planck System: 2D Case}

\author{Zilong Song, Xiulei Cao, Huaxiong Huang}
\affiliation{Department of Mathematics and Statistics, York University and Fields Institute for Research in Mathematical Sciences, Toronto, Ontario, Canada}

\begin{abstract}
The Poisson-Nernst-Planck (PNP) system is a standard model for describing ion transport. In many applications, e.g., ions in biological tissues, the presence of thin boundary layers poses both modelling and computational challenges. In a previous paper, we derived simplified electro-neutral (EN) models in one dimensional space where the thin boundary layers are replaced by effective boundary conditions. In this paper, we extend our analysis to the two dimensional case where the EN model enjoys even greater advantages. First of all, it is much cheaper to solve the EN models numerically. Secondly, EN models are easier to deal with compared with the original PNP system, therefore it is also easier to derive macroscopic models for cellular structures using EN models. The multi-ion case with general boundary is considered, for a variety of boundary conditions including either Dirichlet or flux boundary conditions. Using systematic asymptotic analysis, we derive a variety of effective boundary conditions directly applicable to the EN system for the bulk region. To validate the EN models, numerical computations are carried out for both the EN and original PNP system, including the propagation of action potential for both myelinated and unmyelinated axons. Our results show that solving the EN models is much more efficient than the original PNP system.
\end{abstract}

\maketitle

\section{Introduction}
Ion transport plays a critical role in normal biological functions, and in many cases, excessive charges accumulate next to cell membranes and form thin boundary layers (BLs). These boundary layers constantly adapt to the in- and ef-fluxes of ions through pores formed by proteins embedded in cell membranes, affecting membrane potential and therefore cellular functions. When the overall flux is negligible, and these changes in the BLs occur over a time scale shorter than that of the normal biological function, one can approximate the charge accumulation in the BL by an effective capacitor. On the other hand, when the overall flux is not small, ignoring these changes lead to inconsistency in the electro-neutral status of ionic solution away from these thin layers. In Rubinstein's book \cite{rubinstein1990}, effective boundary condition were derived so that BLs can be ignored when the main interest of the investigation is focused on the evolution of bulk ionic concentration, under the Dirichlet conditions for ions. Extension to other boundary conditions including flux conditions was given in our previous work \cite{song2018} for problem in one dimensional space.

The Poisson-Nernst-Planck (PNP) system is a mathematical model that describes the ion transport under the influence of both an ionic concentration gradient and an electric field.  It is essentially a system coupling diffusion and electrostatics, and the nonlinearity comes from the drift effect of electric field on ions. Such a system and its variants have extensive and successful applications in biological systems, particularly ion channels in cell membrane \cite{bob2001,chun2012}. It has also been applied to many industrial fields, such as the semiconductor devices~\cite{markowich2013} and the detection of poisonous lead by ion-selective electrode~\cite{lead2013}.

When applying to the biological systems, the PNP system will possess a small dimensionless parameter. Such a small parameter leads to the presence of BL near the boundary of concerned domain, often called Debye/double layer in literature. For many decades, research efforts have been devoted to BL analysis of PNP systems. For example, singular perturbation analysis of PNP system has been carried out for narrow ion channels with certain geometric structure \cite{bob2008,singer2009}. Geometric singular perturbation approach has been developed to investigate the existence and uniqueness of solutions in stationary PNP system \cite{liu2009,liu2015} as well as the effects of permanent charge and ion size \cite{liu2013,liu2015b}. For a general steady state case, Wang et al. \cite{xiangsheng2014} have managed to reduce the asymptotic solutions to a single scalar transcendental equation.

Generally speaking, in BL analysis, the solution of PNP system consists of two parts, the BL solution near boundary and the bulk solution in interior region of domain. The two solutions are connected by some matching conditions. In one-dimensional (1D) cases, some matching/continuity condition has been proposed, e.g., the continuity of electro-chemical potential in \cite{rubinstein1990}. This has been successfully applied to the study of steady states of 1D systems, showing the existence of multiple steady states with piecewise constant fixed charge \cite{rubinstein1987}. In a previous paper \cite{song2018}, we have conducted a systematic BL study for the 1D dynamical PNP system, and have derived various effective boundary conditions. We have also managed to bring back some high-order contributions into such effective conditions, which are not negligible in most biological applications. However, most practical cases are 2D or 3D, and we will extend the study to the 2D case in this paper (3D is a straightforward generilization). These conditions replace the BL region and have potential applications for deriving macroscopic models \cite{huang2011} for bulk region in complicated structures. For example,  some macro-equations are often derived in bulk region for the lens circulation \cite{lens1985,lens2013}, by taking into account the fluxes through membranes but ignoring the BL (so the fluxes calculated there might not be accurate).

Besides the BL analysis, many conservative numerical schemes have been developed for PNP systems, such as finite element method \cite{gao2017}, finite-difference scheme \cite{liu2014}, finite volume method \cite{chainais2003,chainais2003b,cao2018}, in one- and higher-dimensional spaces~\cite{lu2010,mirzadeh2014}. It is well-known that one challenge of computation of PNP is how to accurately capture the BL. Since the functions change rapidly in BL, one needs more mesh points in BL than in the bulk region to attain certain accuracy, requiring some techniques like adaptive refined mesh and moving mesh \cite{weizhanghuang1996,tang2003}. This puts more computational cost, especially when there are many BLs in a complicated system. One attractive idea is to derive effective conditions at the boundary to avoid the need of resolving the BL, so that computation is only needed for the bulk region. This becomes extremely important in the 2D case, and is the other motivation of the current work. Our Electro-neutral (EN) model with effective boundary conditions can be solved with much less computational power, compared to the original PNP system. This will be demonstrated in many numerical examples, in particular the propagation of action potential along axon.

The rest of the paper is structured as follows. In Section II, we present the EN theories. First, to illustrate the ideas, we will first study the two-ion case with a circular boundary. Then, it is easily generalized to multi-ion case with general boundaries. In Section III, these effective boundary conditions are validated by some numerical examples. In Section IV, we study one specific biological application, i.e., the propagation of action potential along an axon.  Our EN model, together with effective interface conditions, is very efficient to capture the propagation of action potential. Finally conclusions and discussion of future directions are given in Section V.

\section{The 2D electro-neutral theories}

In this section, we investigate the 2D dynamical PNP system, and derive electro-neutral (EN) systems with various effective boundary conditions.  The 2D domain is set to be $\Omega$ with boundary $\Gamma = \partial \Omega$. First, to illustrate the main ideas, we will study the two-ion case with valencies $\pm 1$ and with a circular boundary $\Gamma$. Then, it is easily generalized to multi-ion case with general boundaries.

Now we briefly recall the 2D dynamical PNP system and introduce some assumptions for deriving EN systems. Suppose there are $n$ ion species, and let $p_i$ be the ion concentrations and $\psi$ be the electric potential. In $\Omega$, the dimensional PNP system is given by
\begin{equation}
\label{eq1}
\begin{aligned}
&- \epsilon_0 \epsilon_r \Delta \psi  = e_0 N_A\sum_{i=1}^n z_i p_i,\\
&\partial_t p_i = - \nabla \cdot \mathbf{J}_{p_i}= D_i \nabla \cdot ( \nabla p_i + \frac{e_0}{k_B T} z_i p_i  \nabla\psi),
\end{aligned}
\end{equation}
where $i=1,..,n$. The first equation is the electrostatic Poisson equation for $\psi(\mathbf{x},t)$ ($\mathbf{x} \in \Omega$), and the second (Nernst-Planck) equation describes the ion transport for each ion species $p_i(\mathbf{x},t)$ ($i=1,..,n$). The quantity $\mathbf{J}_{p_i}$ is the associated flux vector for $p_i$, and $D_i$ is the diffusion constant. The flux consists of two parts, the linear part due to ionic concentration gradient and the nonlinear part from the drift effect of electric filed. Other parameters are vacuum permittivity $\epsilon_0$, relative permittivity $\epsilon_r$, elementary charge $e_0$, Avogadro constant $N_A$, Boltzmann constant $k_B$ and absolute temperature $T$.

In the following, we will consider the dimensionless/normlized version of the above PNP system, see Section IV for details of non-dimensionalization process. We still adopt the same notations, and the PNP system for dimensionless quantities $p,n, \psi$ in the normalized domain $\Omega$ is given by
\begin{equation}
\label{eq2}
\begin{aligned}
&- \epsilon^2 \Delta \psi  = \sum_{i=1}^n z_i p_i,\\
&\partial_t p_i = - \nabla \cdot \mathbf{J}_{p_i}= D_i \nabla \cdot ( \nabla p_i + z_i p_i  \nabla\psi),
\end{aligned}
\end{equation}
where $i=1,..,n$, and $D_i$ are some dimensionless diffusion constants. Here, $\epsilon \ll 1$ is a dimensionless small parameter and defined by
\begin{equation}
\label{eq3}
\begin{aligned}
& \epsilon = \sqrt{\frac{\epsilon_0 \epsilon_r k_B T}{e^2 N_A \tilde{c} L^2}},
\end{aligned}
\end{equation}
where {$\tilde{c}$} is some typical ion concentration and $L$ is some typical length of domain. This system is accompanied by some initial conditions for $p_i$ and some suitable boundary conditions for both $\psi$ and  $p_i$. For example, we may propose either Dirichlet condition or flux condition for each ion species $p_i$. Initial effect is not considered in this work, and we mainly limit ourselves to the case when BL is already present or gradually appears.

As in the 1D case \cite{song2018}, we assume that local electro-neutrality (LEN) condition in bulk region is satisfied, and moreover near global electro-neutrality (NGEN) condition is satisfied, i.e., there is only at most $O(\epsilon)$ unbalanced charge. The second assumption essentially puts some restriction on the boundary conditions, see Remark 3 in later sections. These conditions can be justified in many biological applications, for example in the neuronal axon \cite{hodgkin1990}. It is then natural to assume in the bulk region all the functions concerned and their derivatives are $O(1)$, i.e.,
\begin{equation}
\label{eq4}
\begin{aligned}
&\psi,  \nabla \psi,... \sim O(1), \quad p_i, \partial_{t}p_i,\nabla p_i,... \sim O(1).
\end{aligned}
\end{equation}
In the next subsections, we will derive the EN systems and associated effective boundary conditions based these assumptions.

\subsection{Two-ion case with circular boundary}

In this subsection, we investigate the typical case of two ions with valences $\pm1$, and with the circular boundary. In the following of this subsection, polar coordinates $(r, \theta)$ will be adopted, and we denote cation as $p_1(\mathbf{x},t)= p(r,\theta,t)$ with valency $z_1 =1$ and anion as $p_2(\mathbf{x},t)= n(r,\theta,t)$ with valency $z_2 = -1$. Similarly we write $\psi(\mathbf{x},t) = \psi(r,\theta,t)$.  The fluxes in normal direction ($r$-direction) and circumferential direction ($\theta$-direction) are given by
\begin{equation}
\label{eq5}
\begin{aligned}
& J_p^r = - \left( \frac{\partial p}{\partial r} +  p \frac{\partial \psi  }{\partial r}\right), \quad J_p^\theta  = - \frac{1}{r}\left( \frac{\partial p}{\partial \theta} +  p \frac{\partial \psi  }{\partial \theta}\right),\\
& J_n^r = - \left( \frac{\partial n}{\partial r} - n \frac{\partial \psi  }{\partial r}\right), \quad J_n^\theta  = - \frac{1}{r}\left( \frac{\partial n}{\partial \theta} -n  \frac{\partial \psi  }{\partial \theta}\right).
\end{aligned}
\end{equation}
Then, the original system (\ref{eq2}) is written as 
\begin{equation}
\label{eq6}
\begin{aligned}
&- \epsilon^2 \left(\frac{\partial^2 \psi }{\partial r^2} + \frac{1}{r} \frac{\partial \psi}{\partial r} + \frac{1}{r^2} \frac{\partial^2 \psi}{\partial \theta^2}\right)  = p-n,\\
&- \partial_t p = \frac{\partial J^r_p}{\partial r} + \frac{1}{r} \frac{\partial J^\theta_p}{\partial \theta} + \frac{1}{r} J_p^r,\\
&- \partial_t n =  \frac{\partial J^r_n}{\partial r} + \frac{1}{r} \frac{\partial J^\theta_n}{\partial \theta} + \frac{1}{r} J_n^r,\\ 
\end{aligned}
\end{equation}

Based on the previous assumptions, we obtain approximately the EN condition $p \approx n$ from the first equations in (\ref{eq6}) and more precisely we write
\begin{equation}
\label{eq7}
\begin{aligned}
& p(r,\theta,t;\epsilon) = c(r,\theta,t;\epsilon) +O(\epsilon^2),\\
& n(r,\theta,t ;\epsilon) = c(r,\theta,t;\epsilon) + O(\epsilon^2),\\
&\psi(r,\theta,t;\epsilon) = \phi(r,\theta,t;\epsilon) +O (\epsilon^2),
\end{aligned}
\end{equation}
where $c$ and $\phi$ may depend on $\epsilon$ due to boundary conditions, in other words $c$ and $\phi$ can contain $O(\epsilon)$ terms if boundary conditions have such terms. Thus, the reduced EN system would be 
\begin{equation}
\label{eq8}
\begin{aligned}
&\partial_t c = - \nabla \cdot \mathbf{J}_{c}^+ = \nabla \cdot ( \nabla c+ c \nabla \phi),\\
&\partial_t c = - \nabla \cdot \mathbf{J}_{c}^-=  \nabla \cdot ( \nabla c-  c \nabla \phi),
\end{aligned}
\end{equation}
with remainder $O(\epsilon^2)$, and it is equivalent to 
\begin{equation}
\label{eq8_1}
\begin{aligned}
&\partial_t c = \Delta c, \quad \nabla \cdot ( c \nabla \phi) =0.
\end{aligned}
\end{equation}
In polar coordinates, we write (\ref{eq8}) as
\begin{equation}
\label{eq9}
\begin{aligned}
- \partial_t c =  \frac{\partial J^{r,\pm}_c}{\partial r} + \frac{1}{r} \frac{\partial J^{\theta,\pm}_c}{\partial \theta} + \frac{1}{r} J_c^{r,\pm},\\ 
\end{aligned}
\end{equation}
where
\begin{equation}
\label{eq10}
\begin{aligned}
& J_{c}^{r,\pm}= - \left( \frac{\partial c}{\partial r} \pm  c \frac{\partial \phi}{\partial r}\right),\quad J_{c}^{\theta,\pm}= - \frac{1}{r} \left( \frac{\partial c}{\partial \theta} \pm  c \frac{\partial \phi}{\partial \theta}\right).
\end{aligned}
\end{equation}
Then the objective is to find effective boundary conditions for the EN system, based on the exact boundary conditions of original system.

\subsubsection{Dirichlet boundary condition}

Suppose that the boundary $\Gamma$ is a circle with radius $r_0\sim O(1)$ and the domain $\Omega$ is the region inside the circle. Consider the Dirichlet boundary conditions at $r=r_0$
\begin{equation}
\label{eq11}
\begin{aligned}
& \psi (r_0,\theta,t) = \psi_0(\theta,t),\quad p(r_0,\theta,t)= p_0(\theta,t),\\
& n(r_0,\theta,t)=n_0(\theta,t).
\end{aligned}
\end{equation}
Hereafter, subscript 0 represents quantities at $r=r_0$. 

Under the assumptions of LEN and NGEN and from some previous steady state analysis \cite{rubinstein1990,Lee2010}, we expect a BL with thickness $O(\epsilon)$ near the domain boundary $r=r_0$.  In BL, we have
\begin{equation}
\label{eq12}
\begin{aligned}
& \psi,n,p \sim O(1), \quad \partial_t p, \partial_t n \sim O(1),\\
& \partial_\theta p, \partial_\theta n \sim O(1), \quad \mathbf{J}_n, \mathbf{J}_p \sim O(1),\\ 
&\partial_r \psi, \partial_r p, \partial_r n\sim O({1}/{\epsilon}), \quad \partial_{rr} \psi \sim O({1}/{\epsilon^2}),\end{aligned}
\end{equation}
and thus make the transformation
\begin{equation}
\label{eq13}
\begin{aligned}
& \Phi (R,\theta,t) = \psi(r,\theta,t), \quad N(R,\theta,t)= n(r,\theta,t), \\
& P(R,\theta,t) = p(r,\theta,t),\quad R= \frac{r_0-r}{\epsilon},
\end{aligned}
\end{equation}
where all of the new functions $\Phi,P,N$ and their derivatives are assumed to be $O(1)$. In the following the arguments $(\theta,t)$ in functions will be omitted for brevity. With this transformation and the assumptions, the solution in BL is essentially a 1D problem similar to \cite{song2018}, but for the completeness of this work, we also present it below.

With such scaling, the leading order system of equations in BL is 
\begin{equation}
\label{eq14}
\begin{aligned}
&- \partial_{RR}\Phi  = P- N, \\
& \partial_{R} ( \partial_{R}P + P \partial_{R} \Phi)  = O(\epsilon),\\
& \partial_{R} ( \partial_{R}N - N \partial_{R} \Phi) =O(\epsilon).\\
\end{aligned}
\end{equation}
Since fluxes are finite, integrating once gives
\begin{equation}
\label{eq15}
\begin{aligned}
& \partial_{R} P + P \partial_{R}\Phi=  O(\epsilon),\quad
\partial_{R} N - N \partial_{R} \Phi= O(\epsilon).\\
\end{aligned}
\end{equation}
Thus by matching BL solution with bulk solution, the effective leading-order boundary conditions are given by
\begin{equation}
\label{eq16}
\begin{aligned}
 \ln c_0  +  \phi_0   =   \ln p_0 +  \psi_0 +O(\epsilon),\\
  \ln c_0  -  \phi_0   =   \ln n_0 -  \psi_0 +O(\epsilon),
 \end{aligned}
\end{equation}
where $c_0$ and $\phi_0$ are the limit values of bulk solution $c(r)$ and $\phi(r)$ at $r=r_0$. This is often referred to as the continuity of electro-chemical potential \cite{rubinstein1990}.

In fact, we easily get the leading-order BL solutions
\begin{equation}
\label{eq17}
\begin{aligned}
& \Phi (R) =  \phi_0 + 2 \ln \frac{1- e^{-\sqrt{2 c_0 } R } \tanh\left(\frac{\phi_0  - \psi_0}{4}\right)}{1+ e^{-\sqrt{2 c_0 } R } \tanh\left(\frac{\phi_0  - \psi_0}{4}\right)} +O(\epsilon),\\
& P(R) = c_0  \left( \frac{1+  e^{-\sqrt{2c_0}R} \tanh\left(\frac{\phi_0 - \psi_0}{4}\right)}{1- e^{-\sqrt{2c_0 }R} \tanh\left(\frac{\phi_0 - \psi_0}{4}\right)}\right)^2 + O(\epsilon),\\
& N(R) = c_0 \left( \frac{1-  e^{-\sqrt{2c_0}R} \tanh\left(\frac{\phi_0  - \psi_0}{4}\right)}{1+ e^{-\sqrt{2c_0 }R} \tanh\left(\frac{\phi_0 - \psi_0}{4}\right)}\right)^2 + O(\epsilon),
\end{aligned}
\end{equation}
where the constants  $c_0,\phi_0,\psi_0$ are functions of $(\theta,t)$. The composite solutions are given by 
\begin{equation}
\label{eq18}
\begin{aligned}
&p(r) = P(R) + c(r)- c_0 +O(\epsilon),\\
&n(r) = N(R) + c(r)- c_0 +O(\epsilon),\\
&\psi(r) = \Phi(R) + \phi(r) - \phi_0 + O(\epsilon),
\end{aligned}
\end{equation}
which are uniformly valid in the domain $\Omega$. Since in the bulk we have $p(r) =c(r) +O(\epsilon^2)$ by (\ref{eq7}), it is reasonable to expect $p(r) = c(r) +o(\epsilon)$ in some intermediate region $r_0-r\sim O(\epsilon^\alpha)$ with $0<\alpha<1$, say $\alpha=1/2$.

Next, we consider the $O(\epsilon)$ correction term, since we have kept such terms in equations (\ref{eq8}) for $c(r)$. We take cation $p(r)$ for example. The transport equation can be written as
\begin{equation}
\label{eq19}
\begin{aligned}
&  \frac{\partial (rJ^r_p)}{\partial r} = - r\frac{\partial p}{\partial t} -  \frac{\partial J^\theta_p}{\partial \theta}.
\end{aligned}
\end{equation}
In BL with $r= r_0 - \epsilon R$ and $\partial_r =- \frac{1}{\epsilon}  \partial_R$, we have
\begin{equation}
\label{eq20}
\begin{aligned}
&  J^r_p (R)  =  J_{p,0}^{r}(R) +O(\epsilon R),
\end{aligned}
\end{equation}
where $J_{p,0}^{r}$ is some unknown normal flux at the boundary $r=r_0$. Here, we have used the assumption in (\ref{eq12}) and hence the right-hand side of (\ref{eq19}) multiplied by $\epsilon$ has put into $O(\epsilon R)$ in (\ref{eq20}). Then, by definition of $J^r_p$ in (\ref{eq5}) and using the scale (\ref{eq13}), we get
\begin{equation}
\label{eq21}
\begin{aligned}
& \frac{\partial P}{\partial R} + P \frac{\partial \Phi }{\partial R} = \epsilon J_{p,0}^{r} +O(\epsilon^2 R).
\end{aligned}
\end{equation}
From later section (see Proposition 2), we will see that $J_{p,0}^{r}\approx J_{c,0}^{r,+}$, where $J_{c,0}^{r,+}$ is the limit value of $J_{c}^{r,+}$ in (\ref{eq10}) at $r=r_0$. Therefore, dividing by $P$ and integrating, we obtain
\begin{equation}
\label{eq22}
\begin{aligned}
 &\ln (P(R)) +  \Phi (R)  \\
 =&   \ln p_0 +  \psi_0  + \epsilon J_{c,0}^{r,+}  \int_0^R 1/P(z) dz +O(\epsilon^2 R),
\end{aligned}
\end{equation}
where $P(R),\Phi(R)$ on left-hand side contain $O(\epsilon)$ terms, while for $P(z)$ inside the integral we can use the leading order solution (\ref{eq17}). By matching  \cite{bush1992}, let $R= \epsilon^{\alpha-1} s$ (i.e., $r_0 -r=\epsilon^\alpha s$) with $1/2<\alpha<1$, we get
\begin{equation}
\label{eq23}
\begin{aligned}
& P(\epsilon^{\alpha-1} s) = c (r_0 - \epsilon^\alpha s) +o(\epsilon),\\ 
& \Phi(\epsilon^{\alpha-1} s) = \phi (r_0 - \epsilon^\alpha s) +o(\epsilon).
\end{aligned}
\end{equation}
Taking $R= \epsilon^{\alpha-1} s$ in previous relation (\ref{eq22}), we get from left-hand side
\begin{equation}
\label{eq24}
\begin{aligned}
 & \ln (P(R)) +  \Phi (R) \\
 =&   \ln (c_0 ) + \phi_0   - \left(\frac{\partial_r c(r_0)}{c_0 } + \partial_r\phi(r_0)\right) \epsilon^\alpha s +  o(\epsilon),
\end{aligned}
\end{equation}
and from the integral on right-hand side
\begin{equation}
\label{eq25}
\begin{aligned}
&\epsilon \int_0^R 1/P(z) dz \\
& =\frac{\epsilon^\alpha s}{c_0 } + \frac{\sqrt{2} \epsilon}{c_0^{3/2} }  \left(e^{\left({\psi_0 - \phi_0 }\right)/{2}} -1\right)  + o(\epsilon).
\end{aligned}
\end{equation}
In view of the definition $J_{c,0}^+ = -(\partial_r c(r_0)+ c_0 \partial_r \phi(r_0))$, the $\epsilon^\alpha s$ terms automatically cancel each other (which partially verifies the correctness of matching). Then, we are left with
\begin{equation}
\label{eq26}
\begin{aligned}
& \ln c_0  + \phi_0   -  \frac{\sqrt{2}  J_{c,0}^{r,+}  \epsilon}{(c_0 )^{3/2} } \left(e^{({\psi_0 - \phi_0 })/{2}} -1\right) \\
& = \ln p_0 + \psi_0 +  o(\epsilon),
\end{aligned}
\end{equation}
which can be considered as a generalization of continuity of electro-chemical potential, as there is an $O( \epsilon)$ correction term. The other effective boundary condition from the analysis of anion $n(r)$ is similar, and we summarize the results below.

\noindent \textit{\textbf{Proposition 1.}. Suppose the LEN and NGEN conditions are satisfied, and let $\psi_0(\theta,t)$ and $p_0(\theta,t), n_0(\theta,t)$ be the given electric potential and ion concentrations on circular boundary with radius $r_0$ as in (\ref{eq11}) for PNP system (\ref{eq6}), then we have the effective boundary conditions for the EN system (\ref{eq8})
\begin{equation}
\label{eq27}
\begin{aligned}
& \ln c_0  + \phi_0   -  \frac{\sqrt{2}  J_{c,0}^{r,+}  \epsilon}{(c_0 )^{3/2} } \left(e^{({\psi_0 - \phi_0 })/{2}} -1\right) \\
& = \ln p_0 + \psi_0 +  o(\epsilon),\\
& \ln c_0  - \phi_0   -  \frac{\sqrt{2}  J_{c,0}^{r,-}  \epsilon}{(c_0 )^{3/2} } \left(e^{({\phi_0 - \psi_0 })/{2}} -1\right) \\
& = \ln n_0 - \psi_0 +  o(\epsilon),
\end{aligned}
\end{equation}
where $J_{c}^{r,\pm}$ are defined by (\ref{eq10}) and subscript 0 denotes quantities at $r=r_0$.
}

\noindent {\bf Remark 1.}  One can further derive explicit and asymptotically equivalent boundary conditions for $c_0$ and $\phi_0$
\begin{equation}
\label{eq27_1}
\begin{aligned}
c_0 = &\sqrt{p_0 n_0} + \epsilon \frac{n_0^{1/4} - p_0^{1/4}}{\sqrt{2}\sqrt{ p_0 n_0} } \left( n_0^{1/4} J_{c,0}^{r,+} - p_0^{1/4} J_{c,0}^{r,-} \right),\\
\phi_0 =&  \psi_0 + \frac{1}{2} \ln (p_0/n_0) \\
& + \epsilon \frac{n_0^{1/4} - p_0^{1/4}}{\sqrt{2} n_0 p_0} \left( n_0^{1/4} J_{c,0}^{r,+} + p_0^{1/4} J_{c,0}^{r,-} \right),\\
\end{aligned}
\end{equation}
where
\begin{equation}
\label{eq27_2}
\begin{aligned}
J_{c,0}^{r,\pm}&= - \left.\left( \frac{\partial c}{\partial r} \pm  c_0 \frac{\partial \phi}{\partial r}\right)\right|_{r=r_0} \\
&\approx - \left.\left( \frac{\partial c}{\partial r} \pm \sqrt{p_0 n_0} \frac{\partial \phi}{\partial r}\right)\right|_{r=r_0},
\end{aligned}
\end{equation}
which will be used in numerical examples of later sections.

\subsubsection{Flux boundary condition}

In this subsection, we consider the flux boundary conditions on the circular boundary, and more precisely the normal fluxes together with electric potential are given at $r=r_0$
\begin{equation}
\label{eq28}
\begin{aligned}
& J_p^r(r_0,\theta,t)= J_{p,0}^r(\theta,t),\quad J_n^r(r_0,\theta,t)= J_{n,0}^r(\theta,t),\\
&  \psi (r_0,\theta,t) = \psi_0(\theta,t).
\end{aligned}
\end{equation}
The given flux should be restricted such that the NGEN condition is satisfied. So we expect a BL with thickness $O(\epsilon)$ near boundary. And the aim is to propose suitable effective boundary conditions  for the EN system (\ref{eq8}).

We take cation $p(r)$ for example. From equation $(\ref{eq6})_2$ of PNP system, we easily get for some finite $\delta>0$ (say $\delta = r_0/2$)
\begin{equation}
\label{eq29}
\begin{aligned}
 &(r_0-\delta)J^r_p(r_0-\delta)  \\
 =& r_0 J_{p,0}^{r}  -  \int_{r_0}^{r_0-\delta} \left(r \frac{\partial p}{\partial t} + \frac{\partial J^\theta_p}{\partial \theta}\right) dr,
\end{aligned}
\end{equation}
where arguments $(\theta,t)$ are omitted here and in the following derivation. Similarly, from $(\ref{eq8})_1$ of the EN system, we obtain
\begin{equation}
\label{eq30}
\begin{aligned}
 & (r_0-\delta)J^{r,+}_{c}(r_0-\delta)  \\
  =& r_0 J_{c,0}^{r,+}  -  \int_{r_0}^{r_0-\delta} \left( r \frac{\partial c}{\partial t} + \frac{\partial J^{\theta,+}_{c}}{\partial \theta}\right) dr.
\end{aligned}
\end{equation}
Based on assumptions in (\ref{eq12}) and (\ref{eq7}), we get 
\begin{equation}
\label{eq31}
\begin{aligned}
 &J^{r,+}_{c}(r_0-\delta) = J^r_p(r_0-\delta) + O(\epsilon^2).
\end{aligned}
\end{equation}
Then, immediately combining (\ref{eq29}-\ref{eq31}) gives
\begin{equation}
\label{eq32}
\begin{aligned}
 r_0 J_{c,0}^{r,+} &=    r_0 J_{p,0}^{r} \\
  & - \int_{r_0}^{r_0-\delta} r \frac{\partial (p-c)}{\partial t} + \frac{\partial (J^\theta_p - J^{\theta,+}_{c})}{\partial \theta} dr + O(\epsilon^2).
\end{aligned}
\end{equation}

In the following, we shall simplify the integral in above equation. For simplicity, we denote 
 \begin{equation}
 \label{eq33}
\begin{aligned}
\zeta (\theta,t)= \phi_0 (\theta,t)- \psi_0(\theta,t),
 \end{aligned}
\end{equation}
which is often called zeta potential in the electro-chemistry literature \cite{greenbook,kirby2010}.

The first term in the integral of equation (\ref{eq32}) is calculated as
\begin{equation}
\label{eq34}
\begin{aligned}
& \int_{r_0}^{r_0-\delta} r \frac{\partial (p-c)}{\partial t}  dr \\
& = \int_{r_0}^{r_0-\sqrt{\epsilon}} r \frac{\partial (p-c)}{\partial t}  dr + o(\epsilon)\\
&=\int_{r_0}^{r_0-\sqrt{\epsilon}} r_0 \frac{\partial (p-c)}{\partial t}  dr + o(\epsilon) \\
&=-\epsilon \int_{0}^{\infty} r_0 \frac{\partial (P-c_0)}{\partial t}  dR + o(\epsilon)\\
&= -\epsilon r_0 \partial_t  \left( \sqrt{2 c_0 }  (e^{\zeta/2} -1)\right) + o(\epsilon),
\end{aligned}
\end{equation}
where we have used the assumption that $p=c+o(\epsilon)$ for $r_0-r\ge \sqrt{\epsilon}$, and by setting upper limit of integral as $\infty$ only exponentially small terms are neglected. For the second term in  the integral of equation (\ref{eq32}), we first write
\begin{equation}
\label{eq35}
\begin{aligned}
& J^\theta_p - J^{\theta,+}_{c} \\
=& -\frac{1}{r}\left( \frac{\partial (p-c)}{\partial \theta} +  p \frac{\partial \psi  }{\partial \theta} - c \frac{\partial \phi  }{\partial \theta} \right) \\
=& - \frac{1}{r}\left(  \frac{\partial (p-c)}{\partial \theta}  + (p -c)\frac{\partial \phi}{\partial \theta} + p\frac{\partial (\psi - \phi)}{\partial \theta} \right).
\end{aligned}
\end{equation}
Then, similar to (\ref{eq34}), the integrals of first two parts in (\ref{eq35}) are readily found as
\begin{equation}
\label{eq36}
\begin{aligned}
 & -\int_{r_0}^{r_0-\delta}  \frac{1}{r} \frac{\partial (p-c)}{\partial \theta}  dr  =   \frac{\epsilon}{r_0} \partial_\theta  \left( \sqrt{2 c_0 }  (e^{\zeta/2 } -1)\right) + o(\epsilon),\\
& - \int_{r_0}^{r_0-\delta}  \frac{1}{r} (p -c)\frac{\partial \phi}{\partial \theta}  dr  =  \epsilon \frac{\sqrt{2 c_0 } }{r_0}  \left(   e^{\zeta/2} -1\right) \frac{\partial \phi_0}{\partial \theta} + o(\epsilon).
 \end{aligned}
\end{equation}
For the third part in (\ref{eq35}), by using the explicit solutions (\ref{eq17}), we get
\begin{equation}
\label{eq37}
\begin{aligned}
 & -\int_{r_0}^{r_0-\delta} \frac{p}{r} \frac{\partial (\psi - \phi)}{\partial \theta}  dr \\
 &=- \int_{r_0}^{r_0-\sqrt{\epsilon}} \frac{p}{r_0} \frac{\partial (\psi - \phi)}{\partial \theta}  dr +o(\epsilon)\\
 &= \frac{\epsilon}{r_0}  \int_{0}^{\infty} P(R) \frac{\partial (\Phi - \phi_0)}{\partial \theta}  dR +o(\epsilon)\\
 &=- \frac{\epsilon}{r_0}  \left\{ \sqrt{2 c_0 }  \partial_\theta(e^{\zeta/2})  - (e^{\zeta/2} -1) \partial_\theta \left( \sqrt{2 c_0 }\right)   \right\}+o(\epsilon).
 \end{aligned}
\end{equation}
Combining above formulas in (\ref{eq35}-\ref{eq37}), we obtain
 \begin{equation}
 \label{eq38}
\begin{aligned}
& \int_{r_0}^{r_0-\delta} (J^\theta_p - J^{\theta,+}_{c}) dr \\
& = \frac{\epsilon}{r_0} \left(   e^{\zeta/2} -1\right)  \left[    2 \partial_\theta  \left( \sqrt{2 c_0 } \right) + \sqrt{2 c_0} \partial_\theta ( \phi_0)  \right]  +o(\epsilon).
 \end{aligned}
\end{equation}
Finally from (\ref{eq32},\ref{eq34},\ref{eq38}), the effective boundary condition for bulk flux is given by 
 \begin{equation}
 \label{eq39}
\begin{aligned}
 J_{c,0}^{r,+} =  & J_{p,0}^{r} + \epsilon \partial_t  \left( \sqrt{2 c_0 }  (e^{\zeta/2} -1)\right) \\
 & - \frac{\epsilon}{r_0^2} \partial_\theta \left\{   \sqrt{2 c_0} (e^{\zeta/2} -1)  \partial_\theta\left[ \ln (c_0) + \phi_0 \right] \right\}  + o(\epsilon).
 \end{aligned}
\end{equation}
Likewise, the other effective boundary conditions corresponding to the anion $n(r)$ can be obtained, and we summarize the results below.

\noindent \textit{\textbf{Proposition 2.} Suppose the LEN and NGEN conditions are satisfied, and let $\psi_0(\theta,t)$ and $J_{p,0}^r(\theta,t),  J_{n,0}^r(\theta,t)$ be the given electric potential and ion normal fluxes on circular boundary with $r=r_0$ as in (\ref{eq28}) for PNP system (\ref{eq6}), then we have the effective boundary conditions for the EN system (\ref{eq8}) 
\begin{equation}
\label{eq40}
\begin{aligned}
J_{c,0}^{r,+} &= J_{p,0}^{r} + \epsilon \partial_t  \left( \sqrt{2 c_0 }  (e^{\zeta/2} -1)\right) \\
& - \frac{\epsilon}{r_0^2} \partial_\theta \left\{   \sqrt{2 c_0} (e^{\zeta/2} -1)  \partial_\theta\left[ \ln c_0 + \phi_0 \right] \right\}  + o(\epsilon),\\
J_{c,0}^{r,-} &= J_{n,0}^{r} + \epsilon \partial_t  \left( \sqrt{2 c_0 }  (e^{-\zeta/2} -1)\right) \\
& - \frac{\epsilon}{r_0^2} \partial_\theta \left\{   \sqrt{2 c_0} (e^{-\zeta/2} -1)  \partial_\theta\left[ \ln c_0 - \phi_0 \right] \right\}  + o(\epsilon),
 \end{aligned}
\end{equation}
where $J_c^{r,\pm}$ are defined by (\ref{eq10}), $\zeta$ is defined in (\ref{eq33}) and subscript 0 denotes quantities at $r=r_0$.
}

\noindent\textbf{Remark 2.} Keeping the $O(\epsilon)$ terms in (\ref{eq40}) is necessary for two reasons. First,  in bulk equations  (\ref{eq8}) we have assumed an $O(\epsilon^2)$ remainder so it is reasonable and consistent to bring back the $O(\epsilon)$ terms on boundary conditions. Second, neglecting the $O(\epsilon)$ terms is physically incorrect for EN system as the solution would not be unique (e.g., $\phi$ can differ by a constant).  The effective flux conditions incorporate two effects: (i)  the $ \partial_t$ term accounts for the accumulation of ions in BL, like a capacitor, and (ii) the $\partial_\theta$ term represents the spacial variation along the circumferential boundary. Such terms can be essential in many biological applications, as in the example of action potential in later sections.

\noindent \textbf{Remark 3.} In above proposition, the given fluxes $J_{p,0}^r,J_{n,0}^r$ can be either $O(1)$ or $O(\epsilon)$, as long as the NGEN is satisfied.  This means when fluxes are $O(1)$, we should impose some restriction on the fluxes, i.e.,
\begin{equation}
\label{eq41}
\begin{aligned}
\int_0^t \int_\Gamma (J_{p,0}^r- J_{n,0}^r) d\Gamma dt= O(\epsilon),
 \end{aligned}
\end{equation}
which means the total current flowing into the domain is $O(\epsilon)$. In some cases, the flux is not explicitly given, but is related to the concentrations and electric potential by some model. For example, in biological applications there is Hodgkin-Huxley model \cite{hodgkin1990} or GHK flux model \cite{hille2001}, and for electrolyte there are  Chang-Jaffle boundary conditions \cite{chang1952,chang2015,lead2013}. Suppose the boundary condition is in the form $J_{p,0}^r = f(p_0,\psi_0)$, where $f$ is some given function, then we need to replace $J_{p,0}^r$ by $f$ in Proposition 2 and  supplement these effective flux conditions with those conditions in Proposition 1.

\subsection{Multi-ion case with general boundary}

In this subsection, we extend the preceding results for two ion species to the general multi-ion species case, and consider a domain $\Omega$ inside a general 2D boundary $\Gamma$. We assume that $\Gamma$ is smooth without singularities and that the curvature is not too large, say $O(1)$.

We use curvilinear coordinates to represent a region near boundary. The boundary $\Gamma $ is parametrised by a variable $\eta$, and the distance to the boundary along the normal direction is denoted by $\xi$. The tangent vector along $\Gamma$ is defined by
 \begin{equation}
 \label{eq42}
\begin{aligned}
\mathbf{g}_\eta = \frac{d \mathbf{s}}{d \eta} = g(\eta) \mathbf{e}_\eta,
 \end{aligned}
\end{equation}
where $\mathbf{s} (\eta)$ represents the position vector on the boundary. The function $g(\eta)$ is the metric and $g=1$ if $\eta$ is suitably chosen as the arc length variable, and $\mathbf{e}_\eta$ is the tangent unit vector. The unit normal to the boundary is denoted by $\mathbf{e}_\xi$, pointing inward to be consistent with the definition of variable $\xi$. The curvature $\kappa(\eta)$ on the boundary is defined by
\begin{equation}
\label{eq43}
\begin{aligned}
& d\mathbf{e}_\xi =- \kappa(\eta) d\mathbf{s} = -\kappa (\eta) \mathbf{g}_\eta d\eta,\\
&\kappa(\eta) = -\frac{1}{g(\eta) } \frac{d\mathbf{e}_\xi }{d\eta} \cdot  \mathbf{e}_\eta.
\end{aligned}
\end{equation}
In brief summary, the boundary $\Gamma$ is charaterized two quantities $g(\eta)$ and $\kappa(\eta)$.

For a generic point $\mathbf{x} \in \Omega$ near boundary, we have 
\begin{equation}
\label{eq44}
\begin{aligned}
\mathbf{x} &= \mathbf{s} + \xi \mathbf{e}_\xi, \\  
d\mathbf{x} &= (1-\kappa \xi)d\mathbf{s} + \mathbf{e}_\xi d\xi = \tilde{g}(\eta,\xi) \mathbf{e}_\eta d\eta + \mathbf{e}_\xi d\xi,
\end{aligned}
\end{equation}
where 
\begin{equation}
\label{eq45}
\begin{aligned}
\tilde{g}(\eta,\xi) = (1 -\kappa(\eta) \xi) g(\eta).
\end{aligned}
\end{equation}
Note that for a circle with radius $r_0$ in previous subsection, the above quantities degenerate to
\begin{equation}
\label{eq46}
\begin{aligned}
&g(\eta)=r_0,\quad \mathbf{e}_\xi = -\mathbf{e}_r, \quad \kappa(\eta) =1/r_0, \\ 
&\tilde{g} (\eta,\xi) = r_0 -\xi= r.
\end{aligned}
\end{equation}

Suppose there are $n$ species of ions. Recall that the original PNP system for $p_i$ ($i=1,..,n$) and $\psi$ is given by
\begin{equation}
\label{eq47}
\begin{aligned}
&- \epsilon^2 \Delta \psi  = \sum_{i=1}^n z_i p_i,\\
&\partial_t p_i = - \nabla \cdot \mathbf{J}_{p_i}= D_i \nabla \cdot ( \nabla p_i + z_i p_i  \nabla\psi),\\
\end{aligned}
\end{equation}
where $i=1,..,n$, and $D_i$ are some dimensionless diffusion constants. With previous assumptions and EN conditions, we write 
\begin{equation}
\label{eq48}
\begin{aligned}
& p_i = c_i + O(\epsilon^2),\quad \psi= \phi + O(\epsilon^2).
\end{aligned}
\end{equation}
Then the EN system for bulk region is
\begin{equation}
\label{eq49}
\begin{aligned}
&\partial_t c_i = -  \nabla \cdot \mathbf{J}_{c_i}= D_i  \nabla \cdot ( \nabla c_i + z_i c_i  \nabla \phi),
\end{aligned}
\end{equation}
where $i=1,..,n$. Alternatively, by the EN condition $\sum_{i=1}^n z_i c_i=0$, the EN system for $n$ unknowns $c_1,..,c_{n-1},\phi$ can be written as 
\begin{equation}
\label{eq50}
\begin{aligned}
&\partial_t c_i = - \nabla \cdot \mathbf{J}_{c_i}= D_i  \nabla \cdot ( \nabla c_i + z_i c_i  \nabla \phi),\\
&\sum_{i=1}^n z_i D_i  \nabla \cdot ( \nabla c_i + z_i c_i  \nabla \phi) =0,
\end{aligned}
\end{equation}
for $i=1,..,n-1$ and whenever $c_n$ appears we should replace it by $c_n = -\frac{1}{z_n}\sum_{i=1}^{n-1} z_i c_i$.

In the $(\xi,\eta)$ coordinate system, in some region near boundary $\Gamma$,  the two fluxes for PNP system in the normal and tangential directions are given by
\begin{equation}
\label{eq51}
\begin{aligned}
& J_{p_i}^\xi = \mathbf{e}_\xi \cdot \mathbf{J}_{p_i} = - D_i \left( \frac{\partial p_i}{\partial \xi} + z_i p_i \frac{\partial \psi  }{\partial \xi}\right), \\
& J_{p_i}^\eta = \mathbf{e}_\eta \cdot \mathbf{J}_{p_i} = - \frac{D_i}{\tilde{g}}\left( \frac{\partial p_i}{\partial \eta} +  z_i p_i \frac{\partial \psi  }{\partial \eta}\right),
\end{aligned}
\end{equation}
and similarly the fluxes for EN system are defined by 
\begin{equation}
\label{eq52}
\begin{aligned}
& J_{c_i}^\xi =  -D_i \left( \frac{\partial c_i}{\partial \xi} + z_i c_i \frac{\partial \phi  }{\partial \xi}\right), \\ 
& J_{c_i}^\eta = -\frac{D_i}{\tilde{g}}\left( \frac{\partial c_i}{\partial \eta} +  z_i c_i \frac{\partial \phi  }{\partial \eta}\right).
\end{aligned}
\end{equation}

In the first case, on boundary $\Gamma$ or at $\xi=0$, we consider the boundary conditions of the type
\begin{equation}
\label{eq53}
\begin{aligned}
\psi(0,\eta,t) = \psi_0(\eta,t), \quad J_{p_i}^\xi (0,\eta,t) = J_{p_i,0}^\xi (\eta,t),
\end{aligned}
\end{equation}
where $i=1,..,n$, and subscript $0$ is used to denote the values or limits of quantities at $\xi=0$.

\noindent \textit{{\bf Theorem 1.} Suppose LEN and NGEN conditions are satisfied, and let the boundary $\Gamma$ be parametrized by $\eta$ and characterized by metric $g(\eta)$ and curvature $\kappa(\eta)$, which are supposed to be $O(1)$. Let $\psi_0$ and $J_{p_i,0}^\xi$ be the given electric potential and normal fluxes on boundary $\Gamma$ as in  (\ref{eq53}) for PNP system  (\ref{eq47}), then we have the effective boundary conditions for EN system  (\ref{eq50})
 \begin{equation}
 \label{eq54}
\begin{aligned}
 J_{c_i,0}^{\xi}  =  & J_{p_i,0}^{\xi}  - \epsilon \partial_t  F_{i0} + \frac{\epsilon}{g} \partial_\eta \left\{ \frac{D_i}{g} F_{i0} \partial_\eta \mu_{i0} \right\}+ o(\epsilon),
 \end{aligned}
\end{equation}
where subscript 0 denotes quantities on the boundary $\Gamma$ (i.e., at $\xi=0$), and 
\begin{equation}
\label{eq55}
\begin{aligned}
\mu_{i0} &= \ln c_{i0} + z_i \phi_0,\\
F_{i0} &= F_i(c_{10},..,c_{n-1,0},\phi_0-\psi_0) \\
& = \pm \frac{c_{i0}}{\sqrt{2}} \int_1^{e^{\phi_0 -\psi_0}} \frac{u^{z_i} -1}{\sqrt{\sum_{k=1}^n c_{k0} (u^{z_k} -1)}} \frac{du}{u}.
\end{aligned}
\end{equation}
In $F_i$, the $\pm$ are chosen for the cases $\psi_0\le \phi_0$ and $\psi_0\ge \phi_0$ respectively, but $F_i$ is well-defined around $\phi_0= \psi_0$, and if $F_i$ can be integrated out, the expressions from the two cases are the same. 
}

{\bf Proof:} The derivation follows similar lines as Proposition 2, and here we will mention the key steps different from the previous case. Near boundary $\Gamma$, we adopt the scalings 
\begin{equation}
\label{eq56}
\begin{aligned}
\Phi (X) = \psi(\xi), \quad P_i(X) = p_i(\xi),\quad X= \frac{\xi}{\epsilon},
\end{aligned}
\end{equation}
where $i=1,..,n$, and arguments $(\eta,t)$ are omitted hereafter. In the multi-ion case, the previous explicit solutions in (\ref{eq17}) can not be used anymore. Instead,  by the BL analysis, we get
\begin{equation}
\label{eq57}
\begin{aligned}
 -\partial_{XX} \Phi &= \sum_{i=1}^n z_i P_i(X)  + O(\epsilon) \\
 &= \sum_{i=1}^n z_i c_{i0} e^{z_i (\phi_0 - \Phi(X))} + O(\epsilon).
\end{aligned}
\end{equation}
Integrating once gives
\begin{equation}
\label{eq58}
\begin{aligned}
\partial_{X} \Phi = \pm \sqrt{2 \sum_{i=1}^n c_{i0} \left(  e^{z_i (\phi_0 - \Phi(X))} -1 \right)} + O(\epsilon),
\end{aligned}
\end{equation}
where $\pm$ are chosen for the cases $\psi_0\le \phi_0$ and $\psi_0\ge \phi_0$ respectively. 

In terms of the fluxes  (\ref{eq51}), the transport equation $ (\ref{eq47})_2$ for $p_i$ can be written as
\begin{equation}
\label{eq59}
\begin{aligned}
& -  \frac{\partial p_i}{\partial t} = \frac{\partial J^\xi_{p_i}}{\partial \xi} - \frac{\kappa}{1-\kappa \xi}  J_{p_i}^\xi + \frac{1}{\tilde{g}}\frac{\partial J^\eta_{p_i}}{\partial \eta}.
\end{aligned}
\end{equation}
Multiplying the factor $(1-\kappa \xi)$ on both sides and rearranging terms give
\begin{equation}
\label{eq60}
\begin{aligned}
& \frac{\partial }{\partial \xi} \left( (1- {\kappa}\xi)J^\xi_{p_i} \right) = - (1-\kappa \xi) \frac{\partial p_i}{\partial t} - \frac{1}{g} \frac{\partial J_{p_i}^\eta}{\partial \eta}.
\end{aligned}
\end{equation}
Likewise, the transport equation  (\ref{eq49}) for $c_i$ is
\begin{equation}
\label{eq61}
\begin{aligned}
& \frac{\partial }{\partial \xi} \left( (1- {\kappa}\xi)J^\xi_{c_i} \right) = - (1-\kappa \xi) \frac{\partial c_i}{\partial t} - \frac{1}{g} \frac{\partial J_{c_i}^\eta}{\partial \eta}.
\end{aligned}
\end{equation}
Integrating  (\ref{eq60}) and  (\ref{eq61}) from 0 to $\delta$ and using the fact $J^\xi_{c_i}(\delta) = J^\xi_{p_i} (\delta) + O(\epsilon^2)$ in the bulk, we obtain
\begin{equation}
\label{eq62}
\begin{aligned}
& J_{c_i,0}^{\xi}  = J_{p_i,0}^{\xi}  \\
& - \int_0^\delta \left\{ (1-\kappa \xi)  \frac{\partial (p_i -c_i)}{\partial t} + \frac{1}{g} \frac{\partial (J_{p_i}^\eta -J^{\eta}_{c_i})}{\partial \eta} \right\}d\xi \\
& + O(\epsilon^2),
\end{aligned}
\end{equation}
where $\delta>0$ is some typical bulk value. 

Next, we shall simplify the integral in (\ref{eq62}), by using leading order relations in (\ref{eq57},\ref{eq58}).  We get from the first term that
\begin{equation}
\label{eq63}
\begin{aligned}
& \int_0^\delta (1 -\kappa \xi)  \frac{\partial (p_i -c_i)}{\partial t}  d\xi =  \epsilon \partial_t  F_{i0} + o(\epsilon),\\
& F_{i0} =\int_0^\infty \left( P_i(X)- c_{i0} \right) dX \\
& = \pm \frac{c_{i0}}{\sqrt{2}} \int_1^{e^{\phi_0 -\psi_0}} \frac{u^{z_i} -1}{\sqrt{\sum_{k=1}^n c_{k0} (u^{z_k} -1)}} \frac{du}{u},
\end{aligned}
\end{equation}
where we have made use of the assumption that $\kappa $ is $O(1)$ (or at least $\kappa < O(1/\epsilon)$). We have used only leading order solution of $\Phi$ in $F_{i0}$ and the remainder terms have been put to the $o(\epsilon)$ term.  For the second term in integral of (\ref{eq62}), we write the flux difference as
\begin{equation}
\label{eq64}
\begin{aligned}
& J_{p_i}^\eta -J^{\eta}_{c_i} \\
& = - \frac{D_i}{\tilde{g}} \left( \frac{\partial (p_i - c_i)}{\partial \eta}  + z_i (p_i - c_i) \frac{\partial \phi}{\partial \eta}  + z_i p_i \frac{\partial (\psi - \phi)}{\partial \eta}\right)
\end{aligned}
\end{equation}
and integration leads to
\begin{equation}
\label{eq65}
\begin{aligned}
& \int_0^\delta   (J_{p_i}^\eta -J^{\eta}_{c_i}) d\xi \\
& = - \frac{D_i}{g}  \left( \epsilon \partial_\eta F_{i0} +  \epsilon z_i F_i \partial_\eta \phi_0   +  \int_0^\delta z_i  p_i \frac{\partial (\psi - \phi)}{\partial \eta} d\xi\right) \\
& + o(\epsilon),
\end{aligned}
\end{equation}
where the last term is given by
\begin{equation}
\label{eq66}
\begin{aligned}
& \int_0^\delta z_i p_i \frac{\partial (\psi - \phi)}{\partial \eta} d\xi \\
&= \epsilon \int_0^\infty  z_i P_i(X) \frac{\partial (\Phi - \phi_0)}{\partial \eta} dX \\
&= \epsilon  c_{i0} \int_0^\infty  z_i e^{z_i (\phi_0-\Phi)} \frac{\partial (\Phi - \phi_0)}{\partial \eta} dX \\
& = - \epsilon  c_{i0} \partial_\eta \int_0^\infty (e^{z_i (\phi_0-\Phi)}   -1) dX\\
& =  -\epsilon  c_{i0} \partial_\eta \left(\frac{1}{c_{i0}} F_{i0} \right).
\end{aligned}
\end{equation}
Finally, combining equations (\ref{eq62},\ref{eq63},\ref{eq65},\ref{eq66}) gives the result in (\ref{eq54}). It can be shown as in Appendix A of \cite{song2018} that the function $F_i$ is well-defined near $\phi_0= \psi_0$. \qquad $\square$

\noindent {\bf Remark 4.} In the above effective conditions, the $\partial_t $ term plays a role of a nonlinear capacitor and $\partial_\eta$ term accounts for the ion transport in BL along the boundary. Only the metric parameter $g(\eta)$ is present while the curvature does not influence them as long as it is not very large. In coordinate-free form, the $\partial_\eta$ term becomes
\begin{equation}
\label{eq66_1}
\begin{aligned}
 \frac{\epsilon}{g} \partial_\eta \left\{ \frac{D_i}{g} F_{i0} \partial_\eta \mu_{i0} \right\} = \epsilon \nabla_\Gamma \cdot (D_i F_{i0} \nabla_\Gamma \mu_{i0}),
\end{aligned}
\end{equation}
where $\nabla_\Gamma = \frac{1}{g} \partial_\eta$, and this is similar to a term in equation (2.246) of \cite{mori2006} under linearization of $F_{i0}$. In above 2D case, $\nabla_\Gamma $ is a scalar operator, and for the 3D case $\nabla_\Gamma $ will be a vector operator on surface. In 3D case, the above result in Theorem 1 is still valid with the $\partial_\eta$ term replaced by the right-hand side of $(\ref{eq66_1})$ (in 3D the following theorems 2 and 3 will not change). In some special cases, the function $F_{i0}$ can be integrate out with elementary functions, see Appendix \ref{appendix_A} for details. In view of definition (\ref{eq63}), the term $F_{i0}$ accounts for the accumulation of  i-th ion in BL. For the two-ion case, with (\ref{eq46}) and formula (\ref{eqa1}), the above conditions reduce to those in Proposition 2.

Next, on boundary $\Gamma$ (i.e., $\xi=0$), we consider the boundary conditions of the type
\begin{equation}
\label{eq67}
\begin{aligned}
\psi(0,\eta,t) = \psi_0(\eta,t), \quad  p_{i}(0,\eta,t) = p_{i0}(\eta,t),
\end{aligned}
\end{equation}
where $i=1,..,n$. We summarize the results below.

\noindent \textit{\textbf{Theorem 2.} Suppose the assumptions are the same as Theorem 1.  Let $\psi_0(\eta, t)$ and ${p_{i0}}(\eta, t)$ be the given electric potential and ion concentrations on boundary as in (\ref{eq67}) for original PNP system (\ref{eq47}), then for the EN system (\ref{eq50}) we have the effective boundary conditions  
\begin{equation}
\label{eq68}
\begin{aligned}
& \ln c_{i0} +z_i \phi_0 + \frac{\epsilon J_{c_i,0}^z}{D_i} f_{i0} = \ln p_{i0} + z_i \psi_0 + o(\epsilon),\\
\end{aligned}
\end{equation}
where $i=1,..,n$,  subscript 0 denotes quantities at $\xi=0$, and 
\begin{equation}
\label{eq69}
\begin{aligned}
&f_{i0} = f_i(c_{10},..,c_{n-1,0},\phi_0-\psi_0) \\
=& \pm \frac{1}{\sqrt{2} c_{i0} } \int_1^{e^{\phi_0 -\psi_0}} \frac{u^{-z_i} -1}{\sqrt{\sum_{k=1}^n c_{k0} (u^{z_k} -1)}} \frac{du}{u}.
\end{aligned}
\end{equation}
Here, the $\pm$ are chosen for the cases $\psi_0\le \phi_0$ and $\psi_0\ge \phi_0$ respectively, but $f_i$ is well-defined around $\phi_0= \psi_0$, and if $f_i$ can be integrated out, the expressions from the two cases are the same. }

{\bf Proof:} The derivation follows similar lines as Proposition 1. We only need to start with equation (\ref{eq60}) instead of equation (\ref{eq19}). Then with the scale $X=\xi/\epsilon$, we get
\begin{equation}
\label{eq70}
\begin{aligned}
&J_{p_i}^\xi (X) = J_{p_i,0}^\xi (X) + O(\epsilon X).
\end{aligned}
\end{equation}
Then, similar to (\ref{eq22}), one can get
\begin{equation}
\label{eq71}
\begin{aligned}
 &\ln (P_i(X)) + z_i \Phi (X)  \\
 =&   \ln p_{i0} + z_i \psi_{i0}  - \frac{\epsilon J_{c_i,0}^{\xi}}{D_i}  \int_0^X 1/P_i(z) dz +O(\epsilon^2 X).
\end{aligned}
\end{equation}
Finally the term $f_{i0}$ is defined from the above integral by using leading order approximations (\ref{eq57}, \ref{eq58}).  The explicit expression for some special cases are given in Appendix \ref{appendix_A}.
 \qquad $\square$
 
Finally, we will consider a case with Robin boundary conditions for $\psi$, since this is common in modeling a membrane (see Section IV). More precisely, we have 
\begin{equation}
\label{eq72}
\begin{aligned}
& \gamma \partial_\xi \psi(0,\eta, t) = \psi(0,\eta,t) -\tilde{\psi}_0(\eta,t), \\
& J_{p_i}^\xi (0,\eta,t) = J_{p_i,0}^\xi (\eta,t),
\end{aligned}
\end{equation}
where $\gamma$ is a parameter and $\tilde{\psi}_0$ is some given function. In this case, $\psi (0,\eta,t)$ is not known and so we need an additional condition to determine $\psi_0\equiv \psi(0,\eta, t)$ in flux conditions (\ref{eq54}, \ref{eq55}). From the relation (\ref{eq58}) and with $\partial_X = \epsilon \partial_\xi$, we get at the leading order
\begin{equation}
\label{eq73}
\begin{aligned}
& \epsilon \partial_\xi \psi(0) = \pm \sqrt{2 \sum_{i=1}^n c_{i0} \left(  e^{z_i (\phi_0 - \psi_0)} -1 \right)},
\end{aligned}
\end{equation}
where $\pm$ are chosen for the cases $\psi_0\le \phi_0$ and $\psi_0\ge \phi_0$ respectively. Combining with $(\ref{eq72})_1$ leads to the nonlinear condition for $\psi_0$
\begin{equation}
\label{eq74}
\begin{aligned}
\psi_0 -\tilde{\psi}_0= \pm \frac{\gamma}{\epsilon} \sqrt{2 \sum_{i=1}^n c_{i0} \left(  e^{z_i (\phi_0 - \psi_0)} -1 \right)}.
\end{aligned}
\end{equation}
See Appendix \ref{appendix_A} for more explicit formulas in special cases. In the above derivation, we have tacitly assumed that $\gamma \le O(\epsilon)$, so that the remainder is $o(1)$ in (\ref{eq74}). For the case $O(\epsilon)<\gamma \le O(1)$, with the NGEN assumption in this work, some previous results \cite{Lee2010,song2018} and numerical evidence in Section \ref{sec4} show that $\psi_i - \phi_i= o(1)$, which is consistent with (\ref{eq74}). In fact, in BL we have
\begin{equation}
\label{eq75}
\begin{aligned}
& \psi - \phi_0, p_i-c_{i0}=O(\epsilon/\gamma),\\
& \partial_\xi \psi,\partial_\xi p_i= O(1/\gamma),\quad \partial_{\xi\xi} \psi = O(1/(\gamma \epsilon)),...
\end{aligned}
\end{equation}
So with slight modification of the transformation (e.g., $\Phi = \psi- \phi_0$), one can show that the relation still holds at leading order. We summarize the results below.

\noindent \textit{\textbf{Theorem 3.} Suppose the assumptions are the same as Theorem 1.  Let $\tilde{\psi}_0(\eta, t)$, $J_{p_i,0}^\xi(\eta,t)$ and parameter $\gamma$ be the given as in (\ref{eq72}) for original PNP system (\ref{eq47}), then for the EN system (\ref{eq50}) we have the same effective flux conditions (\ref{eq54},\ref{eq55}) as in Theorem 1 except that $\psi_0$ is determined by (\ref{eq74}).
}

As the Robin-type boundary condition often appears in modelling cell membrane, here we brief mention an example relevant to macroscopic models for cellular structures. Suppose $\psi$ and $\tilde{\psi}$ denote the electric potential inside and outside a cell, and on the tissue scale they are almost a constant $\phi_0$ and $\tilde{\phi}_0$ (say, averaged quantities). But they are not constant in the BL near membrane, and connected by condition (\ref{eq72}) on membrane with $\gamma = \epsilon^2/C_m$ where $C_m$ is some dimensionless membrane capacitance (cf. (\ref{eq108}) in Sec \ref{sec4}). The average of each ion concentration in cell may be defined as 
\begin{equation}
\label{eq75_1}
\begin{aligned}
\bar{p}_i &\equiv \frac{1}{V_{cell}} \int_{V_{cell}} p_i dx \\
&=  \frac{1}{V_{cell}} \int_{V_{cell}} c_i dx +  \frac{1}{V_{cell}} \int_{V_{BL}} p_i - c_i dx \\
& = {c}_{i0} + \frac{S_m}{V_{cell}} \epsilon F_{i0}(c_{10},..,c_{n-1,0}, \phi_0-\psi_0),
\end{aligned}
\end{equation}
where $V_{BL}$ is some region containing the BL, $F_{i0}$ is in (\ref{eq63}), $S_m$ is surface area of cell, $c_i$ in cell is also considered a constant $c_{i0}$ on tissue scale and $\psi_0$ is determined by (\ref{eq74}). As estimated in Remark 10 of \cite{song2018}, we have $O(\epsilon)\ll \gamma<O(1)$, and hence variation $ \phi_0-\psi_0$ is small as in (\ref{eq75}). Then we can simplify (\ref{eq74}) and (\ref{eq75_1}) based on small $ \phi_0-\psi_0$. It is easy to show from (\ref{eq63}) and (\ref{eq74}) that
\begin{equation}
\label{eq75_2}
\begin{aligned}
& z_i  F_{i0} =  \frac{z_i^2 c_{i0}}{\sqrt{\sum_{k=1}^n z_k^2 c_{k0}}} (\phi_0 - \psi_0),\\
& \frac{\gamma}{\epsilon} \sqrt{\sum_{k=1}^n z_k^2 c_{k0}} (\phi_0 - \psi_0) = \psi_0 - \tilde{\psi}_0  \approx \phi_0 - \tilde{\phi}_0,
\end{aligned}
\end{equation}
and thus $F_{i0}$ can be expressed by averaged quantities. By a summation, we see that 
\begin{equation}
\label{eq75_1}
\begin{aligned}
\sum_{i=1}^n z_i \bar{p}_i & = \frac{S_m}{V_{cell}} \epsilon \sum_{i=1}^n z_i F_{i0} =  \frac{S_m}{V_{cell}}  \frac{\epsilon^2}{\gamma} (\phi_0 - \tilde{\phi}_0) \\
&=  \frac{S_m}{V_{cell}}  C_m (\phi_0 - \tilde{\phi}_0). 
\end{aligned}
\end{equation}
This means that the averaged quantities $\bar{p}_i$ on whole cell including BL do not satisfy electro-neutrality exactly, but are approximated by a linear capacitor.  This is one explanation that some works \cite{mori2006,mori2015} in literature can use a capacitor to model BL effect in macroscopic models. More details and application to specific situations will be left as future study.

\section{Numerical examples}

In this section, we present some numerical examples, to verify the previous effective boundary conditions and to show the accuracy of the EN system.


\subsection{A steady state problem}  

As a first example to verify the previous effective conditions, we study a steady state problem \cite{rubinstein1990,song2018}, since it can be solved analytically for the EN system. We consider an annulus domain $\Omega$, which is defined by $1\le r\le 2$ in polar coordinates $(r,\theta)$. We consider a 2D steady state case for two ions $p(r,\theta), n(r,\theta)$ with valencies $z_1= +1, z_2 = -1$. The boundary conditions in $(r,\theta)$ coordinates are 
\begin{equation}
\label{eq78}
\begin{aligned}
&p(1,\theta)=n(1,\theta) =1,\quad \psi(1,\theta) = 0,\\
& p(2,\theta) = 1, \quad J_n^r(2,\theta)=0, \quad \psi(2,\theta)= -V.
\end{aligned}
\end{equation}
Due to symmetry, the original PNP system (\ref{eq6}) reduces to a 1D problem
\begin{equation}
\label{eq79}
\begin{aligned}
&- \epsilon^2 \left(\frac{d^2 \psi }{dr^2} + \frac{1}{r} \frac{d \psi}{d r} \right)  = p-n,\\
&r \left( \frac{d p}{d r} +  p \frac{d \psi  }{d r}\right) = -j\\
&  \frac{d n}{\partial r} -n \frac{d \psi  }{\partial r}= 0,
\end{aligned}
\end{equation}
where $j$ is some flux constant. The aim is to determine the current-voltage $j$-$V$ relation. Hereafter, the argument $\theta$ in functions will be omitted. Since it is electro-neutral at $r=1$, there is only a BL near the outer boundary $r=2$. The EN system (\ref{eq9}) is 
\begin{equation}
\label{eq80}
\begin{aligned}
& \frac{d c}{d r} +  c \frac{d\phi  }{dr} = \frac{-j}{r}\\
&  \frac{dc}{dr} -c \frac{d \phi  }{d r}= 0.
\end{aligned}
\end{equation}
With boundary condition $c(1) = 1, \phi(1) =0$, the 1D analytical solution can be obtained 
\begin{equation}
\label{eq81}
\begin{aligned}
& c (r)= 1- \frac{j}{2} \ln(r),\quad \phi(r) = \ln (c(r)).
\end{aligned}
\end{equation}
By leading order condition or continuity of electro-chemical potential at $r=2$ (see (\ref{eq16})), we get
\begin{equation}
\label{eq82}
\begin{aligned}
& j= \frac{2(1 - e^{-V/2})}{\ln 2}.
\end{aligned}
\end{equation}
The present effective condition (\ref{eq27}) implies 
\begin{equation}
\label{eq83}
\begin{aligned}
& 2 \ln \left( 1- \frac{j}{2} \ln 2 \right)  \\
&- 2\epsilon j \left( \frac{\sqrt{2} e^{-V/2}}{(2-j \ln 2)^{2}} - \frac{1}{(2-j \ln 2)^{3/2}} \right)= -V,
\end{aligned}
\end{equation}
where an $O(\epsilon)$ correction is present.

\begin{figure}[h]
\begin{center}
\includegraphics[width=3.1in]{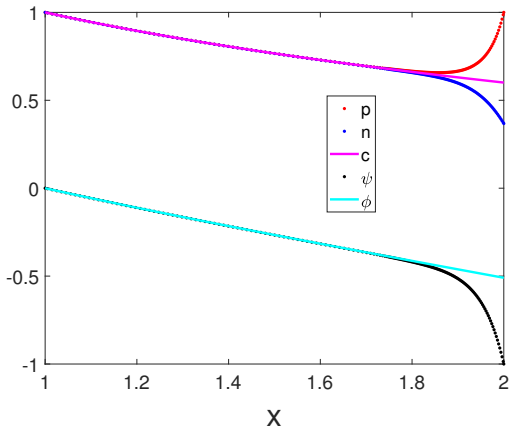}
\caption{\label{figure1} Comparison between analytic bulk solution with numerical solution at $t=20$, with $\epsilon=0.05$. Dots represent the exact solutions of $p,n,\psi$, and solid lines are the approximate solutions of $c,\phi$.}
\end{center}
\end{figure}

In the numerical verification, we use the dynamic system (\ref{eq6}) with boundary conditions (\ref{eq78}) and the following initial conditions at $t=0$,
\begin{equation}
\label{eq84}
\begin{aligned}
p(r,\theta,0) = 1,\quad n(r,\theta,0)=1.
\end{aligned}
\end{equation} 
The solution tends to the steady state solution of (\ref{eq78}) and (\ref{eq79}), and the flux $j$ near the steady state can be found. Finite-volume method with refined mesh near outer boundary $r=2$ is adopted in the numerical simulation, since we require more accuracy for flux $j$. The flux $j$ at time $t=20$ is almost a constant and used as the exact value. With $V=1$ and $\epsilon = 0.1, 0.05, 0.01$, we give the results of flux $j$ using leading order condition (\ref{eq82}) and the present condition (\ref{eq83}) in Table \ref{table1}. It can be seen that the present effective condition produces better results and the $O(\epsilon)$ term is correct. Figure \ref{figure1} shows the good agreement in the bulk region between EN solution (\ref{eq81}) with flux in (\ref{eq83})  and the numerical solution at $t=20$ with $\epsilon=0.05$. In order to show the error of solution with respect to small parameter $\epsilon$, Table \ref{table1} compares the maximum errors of $c(r)$ and $\phi(r)$ by (\ref{eq81}), in the bulk region $[1,1.5]$ with different $\epsilon$.

\begin{table}
\begin{center}
\begin{tabular}{c|c|c|c}
\hline\hline
$\epsilon$ & 0.1 & 0.05& 0.01\\
Leading & 1.1353 \quad &1.1353 \quad &1.1353 \\
Present & 1.1687 & 1.1519 &  1.1386\\
PNP & 1.1718 & 1.1527 & 1.1387\\
\hline\hline
\end{tabular}
\caption{\label{table1}Comparison of flux $j$ with fixed $V=1$ and different $\epsilon$, where ``Leading" and ``Present" are from formulas  (\ref{eq82}) and  (\ref{eq83}), ``PNP" is obtained by solving the dynamic PNP system.}
\end{center}
\end{table}

\begin{table}
\begin{center}
\begin{tabular}{c|c|c|c}
\hline\hline
$\epsilon$ & 0.1 & 0.05& 0.01\\
PNP $|p-n|$ &$4.8232\times10^{-3}$  \quad&$7.3240\times10^{-4}$  \quad&$3.1258\times10^{-5}
$ \\
$|c-p|$ &$2.8585\times10^{-3}$ &$1.4192\times10^{-3}$  &$5.6801\times 10^{-4}$ \\
$|\psi-\phi|$ &$5.2579\times10^{-3} $ \quad&$1.8024\times10^{-3}$  \quad& $5.8205\times 10^{-4}$\\
\hline\hline
\end{tabular}
\caption{\label{table2} Comparison of maximum errors of  $c(r)$ and $\phi(r)$ in the bulk region $r \in [1,1.5]$ with different $\epsilon$, where $p$ and $n$ are from dynamic PNP system, and $c$ and $\phi$ are from (\ref{eq81}) with associated flux $j$ in Table \ref{table1}.}
\end{center}
\end{table}

\subsection{A dynamic problem with Dirichlet conditions} 

Now we consider the circular domain $\Omega$ defined by $r\le1$, and study a dynamic two-ion case with Dirichlet boundary conditions. The original PNP system for $p,n,\psi$ is given by (\ref{eq5},\ref{eq6}), and the boundary conditions are adopted as
\begin{equation}
\label{Eq69}
\begin{aligned}
&\psi (1,\theta,t)= 0, \quad p(1,\theta,t) = 1+ t \sin (|\theta|/2),\\
& n(1,\theta,t) = 1+ t \cos(|\theta|/2), \quad -\pi<\theta\le \pi.
\end{aligned}
\end{equation}
In this example, both $p$ and $n$ increase from 1 as time evolves, but the increased magnitudes are different between $p$ and $n$ for fixed $\theta$, and therefore BL will gradually appear. We take $\epsilon=0.05$ as an illustration, and finite element method with refined mesh near boundary $r=1$ is used to solve this system.

\begin{figure}[h]
\begin{center}
\subfigure[$p$ from PNP system]{\includegraphics[width=3.1in]{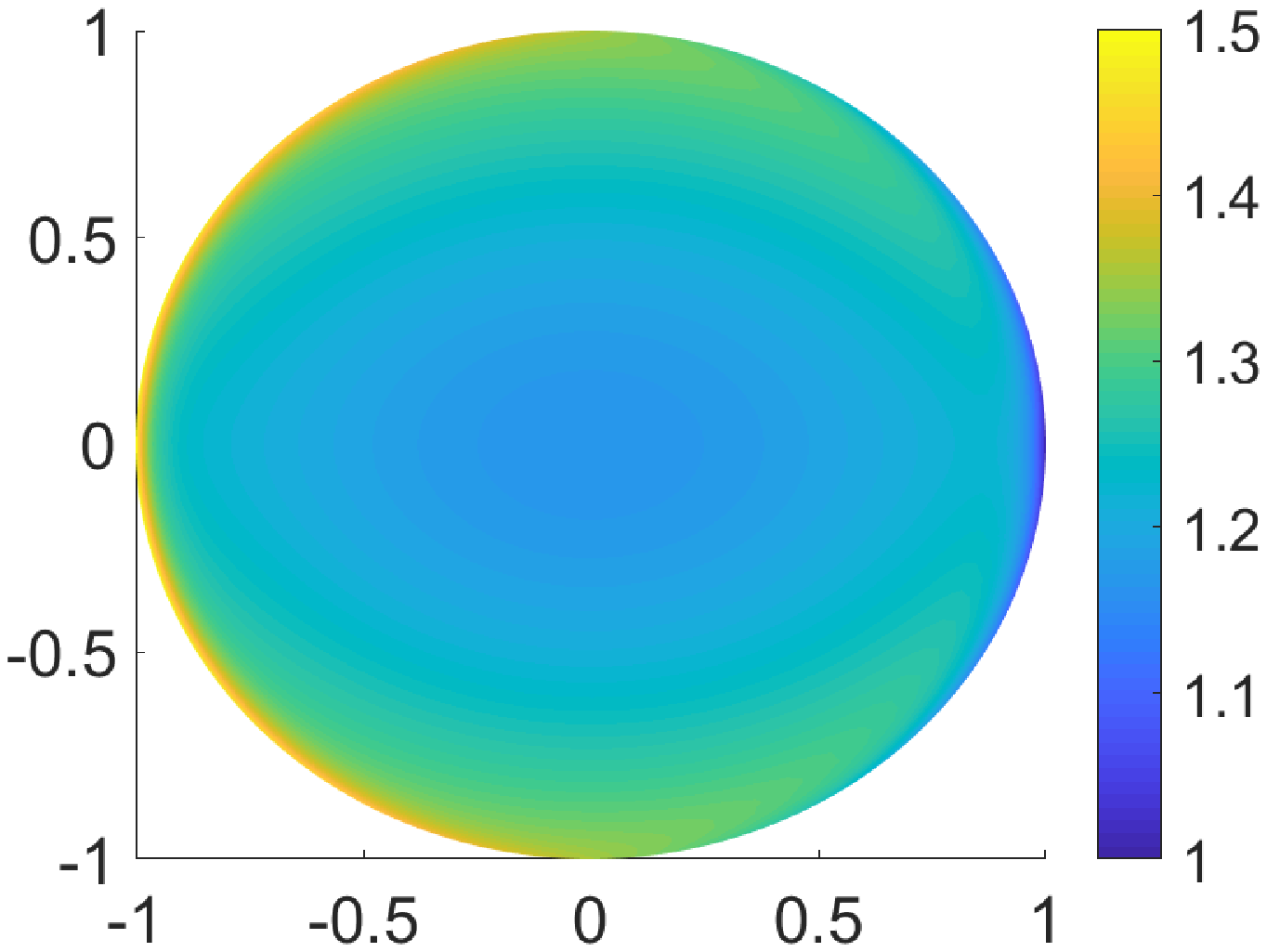}}
\subfigure[$c$ from EN model]{\includegraphics[width=3.1in]{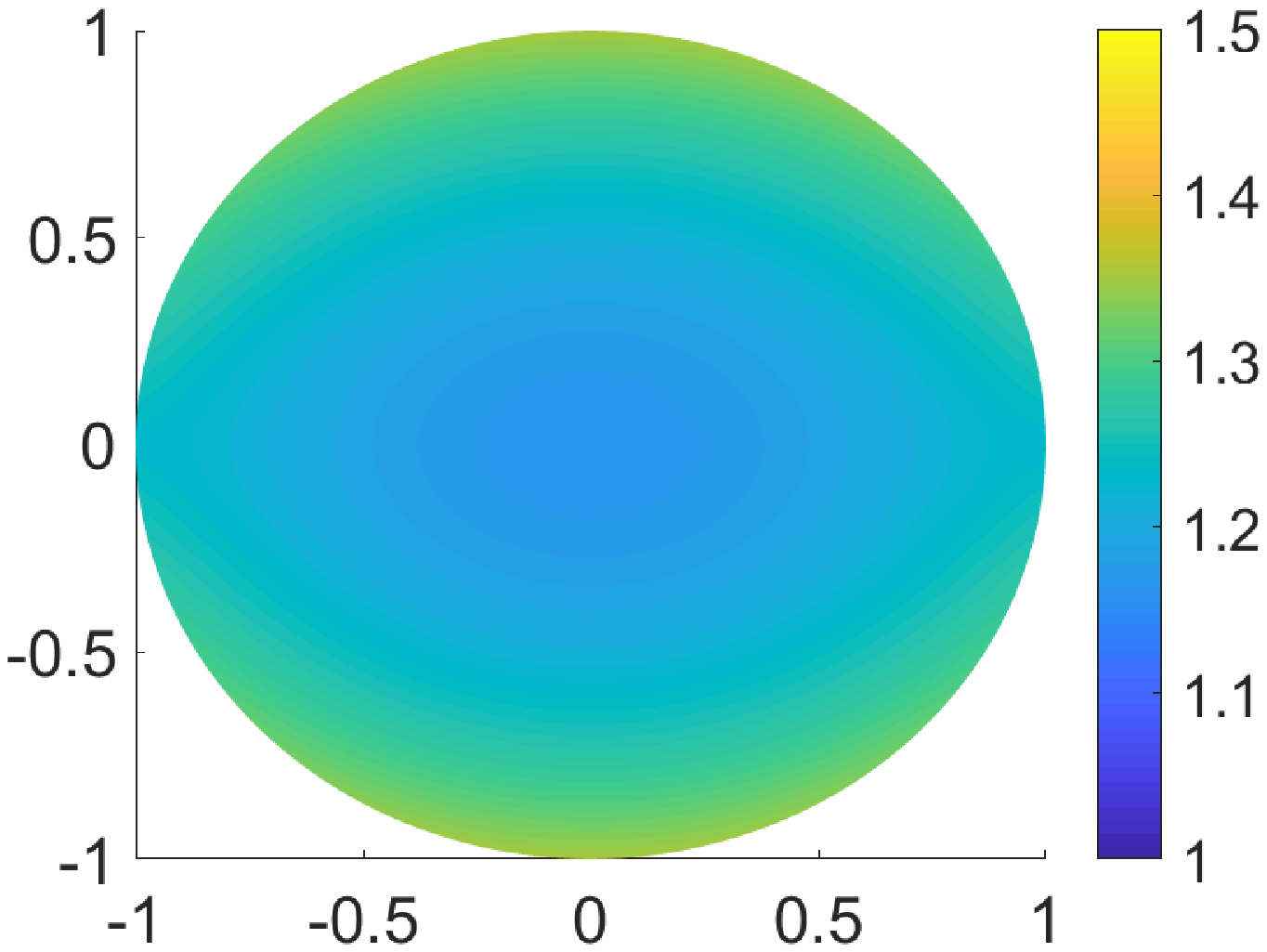}}
\caption{\label{figure2} Comparison of concentrations $p(r,\theta,t)$ from PNP system and $c(r,\theta,t)$ from EN model with present condition (\ref{eq27_1}) at $t=0.5$. }
\end{center}
\end{figure}

\begin{figure}[h]
\begin{center}
\subfigure[$\psi$ from PNP system]{\includegraphics[width=3.1in]{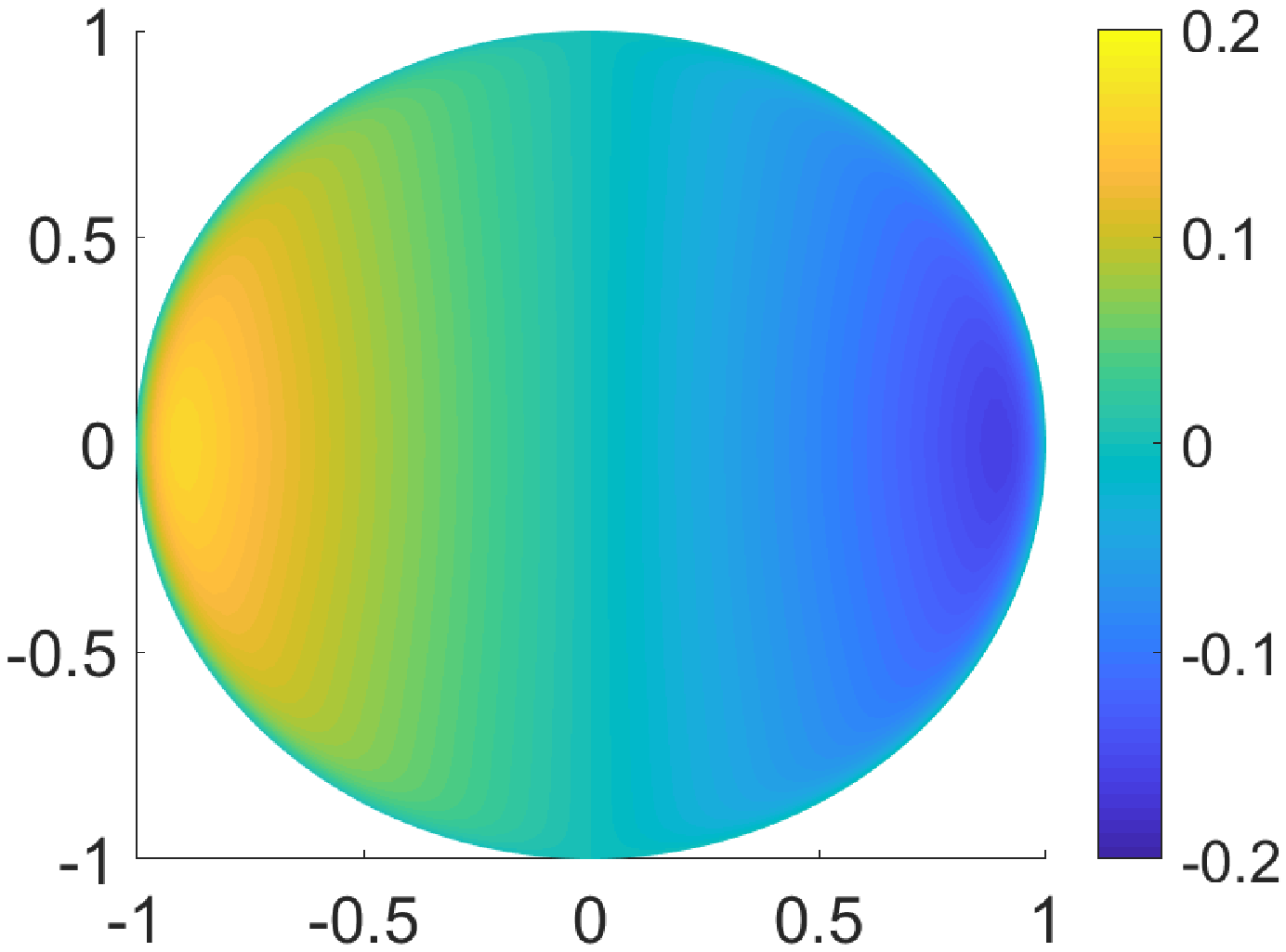}}
\subfigure[$\phi$ from EN model]{\includegraphics[width=3.1in]{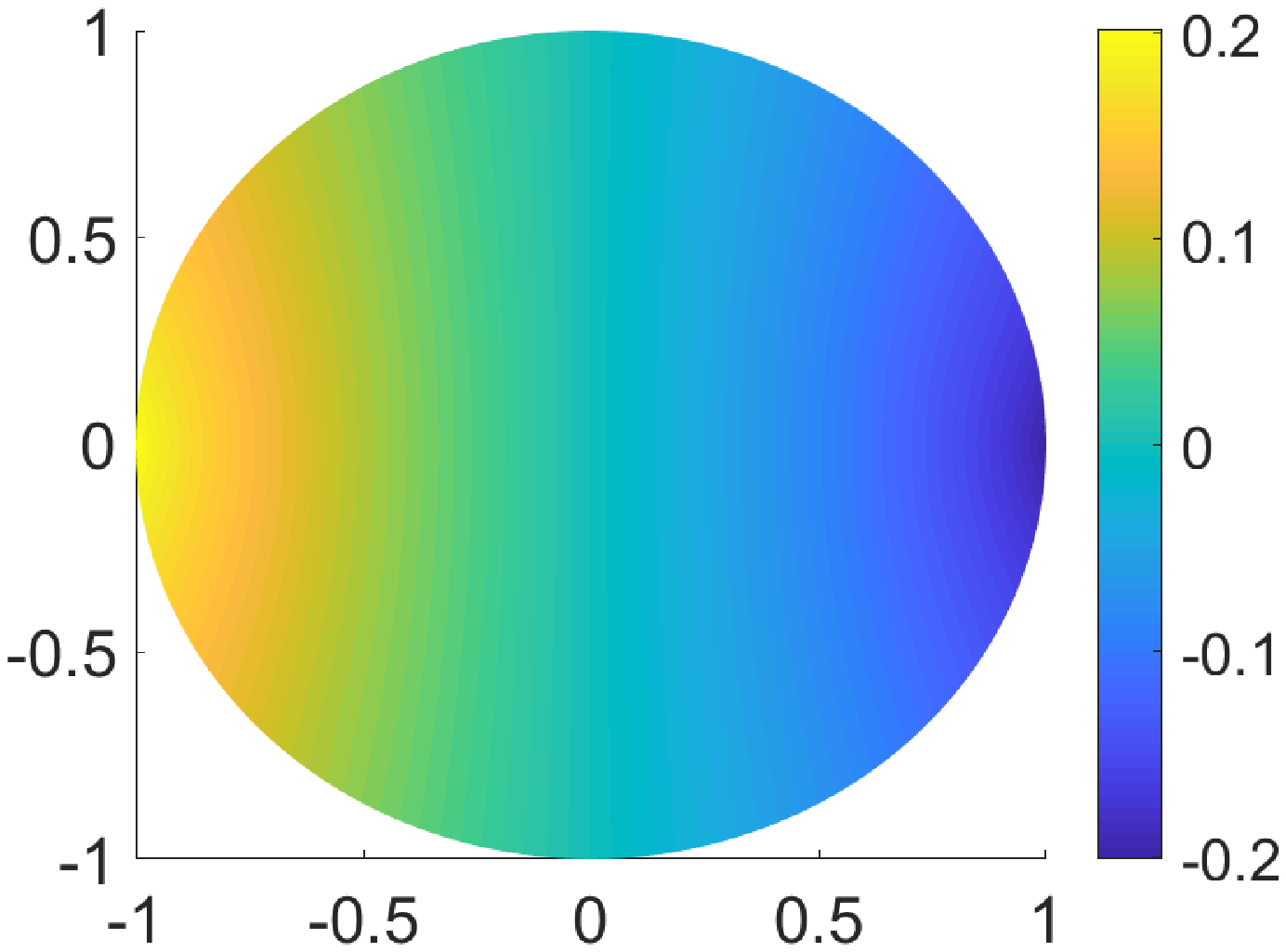}}
\caption{\label{figure3} Comparison of electric potentials $\psi(r,\theta,t)$ from PNP system and $\phi(r,\theta,t)$ from EN model with present condition (\ref{eq27_1}) at $t=0.5$. }
\end{center}
\end{figure}

In this example, the EN system in (\ref{eq9},\ref{eq10}) is solved with effective conditions in  (\ref{eq27_1},\ref{eq27_2}) in Remark 1. More precisely, finite element method (without refined mesh near boundary) is also used in the simulation. We conduct two implementations, (i) with leading order boundary condition 
\begin{equation}
\label{Eq70}
\begin{aligned}
c(1,\theta,t) = &\sqrt{p(1,\theta,t) n (1,\theta,t)},\\
\phi_0(1,\theta,t) =&  \frac{1}{2} \ln (p(1,\theta,t)/n(1,\theta,t)),
\end{aligned}
\end{equation}
and (ii) the high-order boundary condition  (\ref{eq27_1},\ref{eq27_2}) with $O(\epsilon)$ term, where explicit method is used to treat the fluxes $J_{c,0}^{r,\pm}$ (here at $r=1$) by the estimate from previous time step.

By using the numerical results of $p(x,t)$ and $\psi(x,t)$ of the original system as a reference solution, Table \ref{table3} gives the maximum errors of $c(x,t)$ and $\phi(x,t)$ in some bulk region $r\in[0,0.5]$ at $t=0.5$. The results indicate that the accuracy is very good with the effective boundary conditions. Figure \ref{figure2} shows the comparison between $p(r,\theta,t)$ from PNP system and $c(r,\theta,t)$ from EN model, and Figure \ref{figure3} shows the comparison between $\psi(r,\theta,t)$ from PNP system and $\phi(r,\theta,t)$ from EN model with boundary condition (\ref{eq27_1}) at $t=0.5$. They show that the approximate solutions $c(x,t)$ and $\phi(x,t)$ agree very well with exact solutions. Furthermore, the EN system allows for relatively large mesh and time step sizes, and as a result the computational time is greatly reduced. For instance, it takes roughly 4.8 hours to compute the original PNP system up to $t=0.5$ while it takes only 2 minutes for the EN system on the same computer (Processor: 4GHz, i76700K; Memory: 32GB).


\begin{table}
\begin{center}
\begin{tabular}{c|c|c}
\hline\hline
& $|c-p|$ & $|\phi-\psi|$\\
Leading & $4.6304\times10^{-4}$ \quad  &2.7890$\times 10^{-4}$ \\
Present & $3.0312\times 10^{-5}$  &  $1.3641\times 10^{-4}$\\
PNP $|p-n|$ &  3.3183$\times10^{-5}$ &  --\\
\hline\hline
\end{tabular}
\caption{\label{table3}  Maximum error in concentration $c(x,t)$ and potential $\phi(x,t)$ in some bulk region $r\in[0,0.5]$ and $t=0.5$, using leading order condition (\ref{Eq70}) and present condition (\ref{eq27_1}).}
\end{center}
\end{table}

\subsection{A dynamic problem with flux conditions} 

As a second dynamic example, we study the two-ion case in circular domain $\Omega$  with flux conditions. More precisely, we propose at $r=1$,
\begin{equation}
\label{Eq71}
\begin{aligned}
&\psi (1,\theta,t)= 0, \quad J_p^r (1, \theta,t)= 4 \epsilon \sin (\theta), \\
& J_n^r (1, \theta,t)= 2 \epsilon \cos (\theta), \quad -\pi<\theta\le \pi,
\end{aligned}
\end{equation}
where $\epsilon= 0.05$ as in the previous example. In this example, the integral of $J_p^r$ or $J_n^r$ over entire boundary ($\theta$ from $-\pi$ to $\pi$) will be 0, and so the global electro-neutrality is automatically satisfied. 

\begin{figure}[h]
\begin{center}
\subfigure[ $p$ from PNP system]{\includegraphics[width=3.1in]{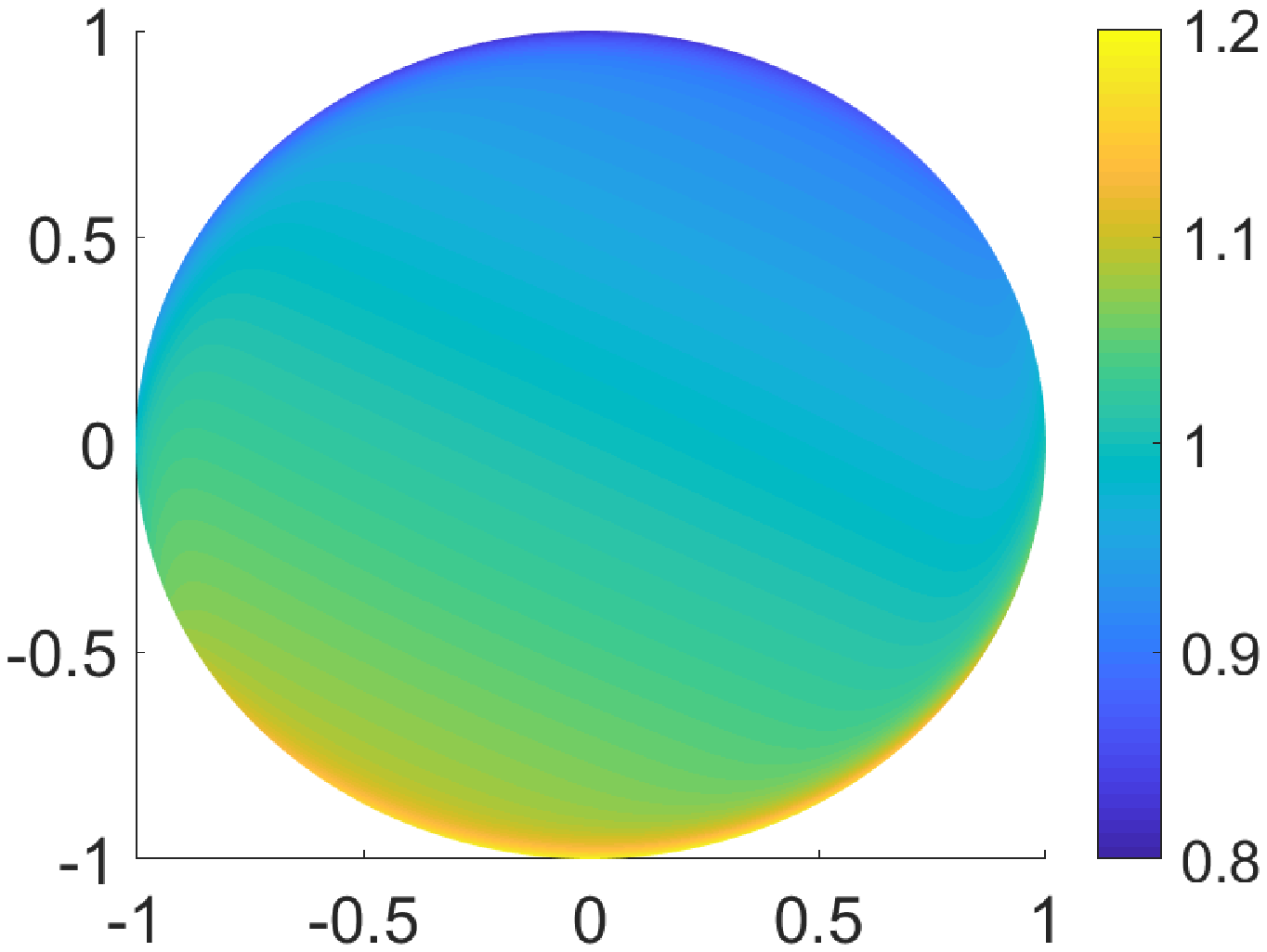}}
\subfigure[ $c$ from EN model]{\includegraphics[width=3.1in]{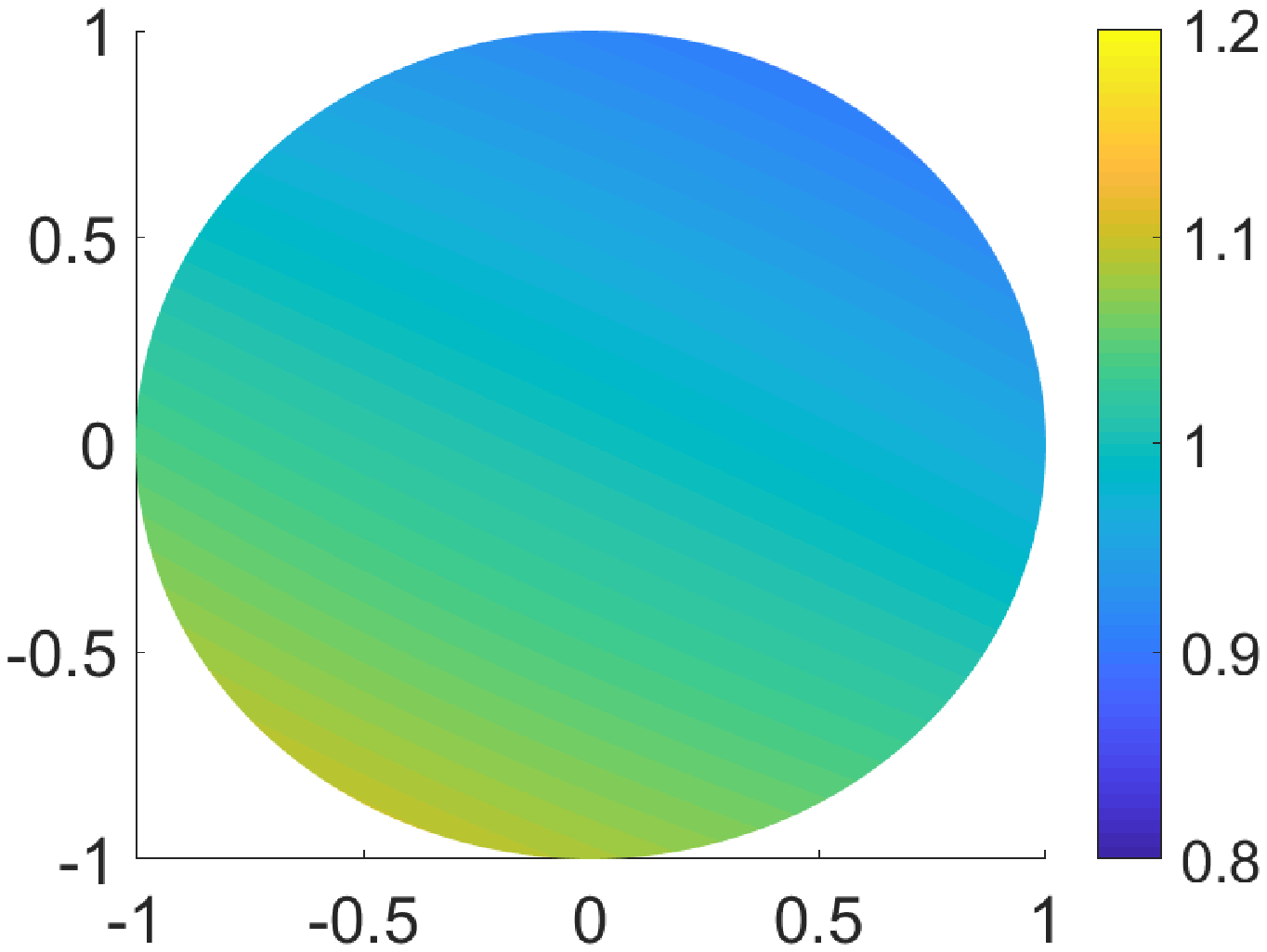}}
\caption{\label{figure4} Comparison of concentrations $p(r,\theta,t)$ from PNP system and $c(r,\theta,t)$ from EN model with present condition (\ref{eq27_1}) at $t=0.5$. }
\end{center}
\end{figure}

\begin{figure}[h]
\begin{center}
\subfigure[$\psi$ from PNP system]{\includegraphics[width=3.1in]{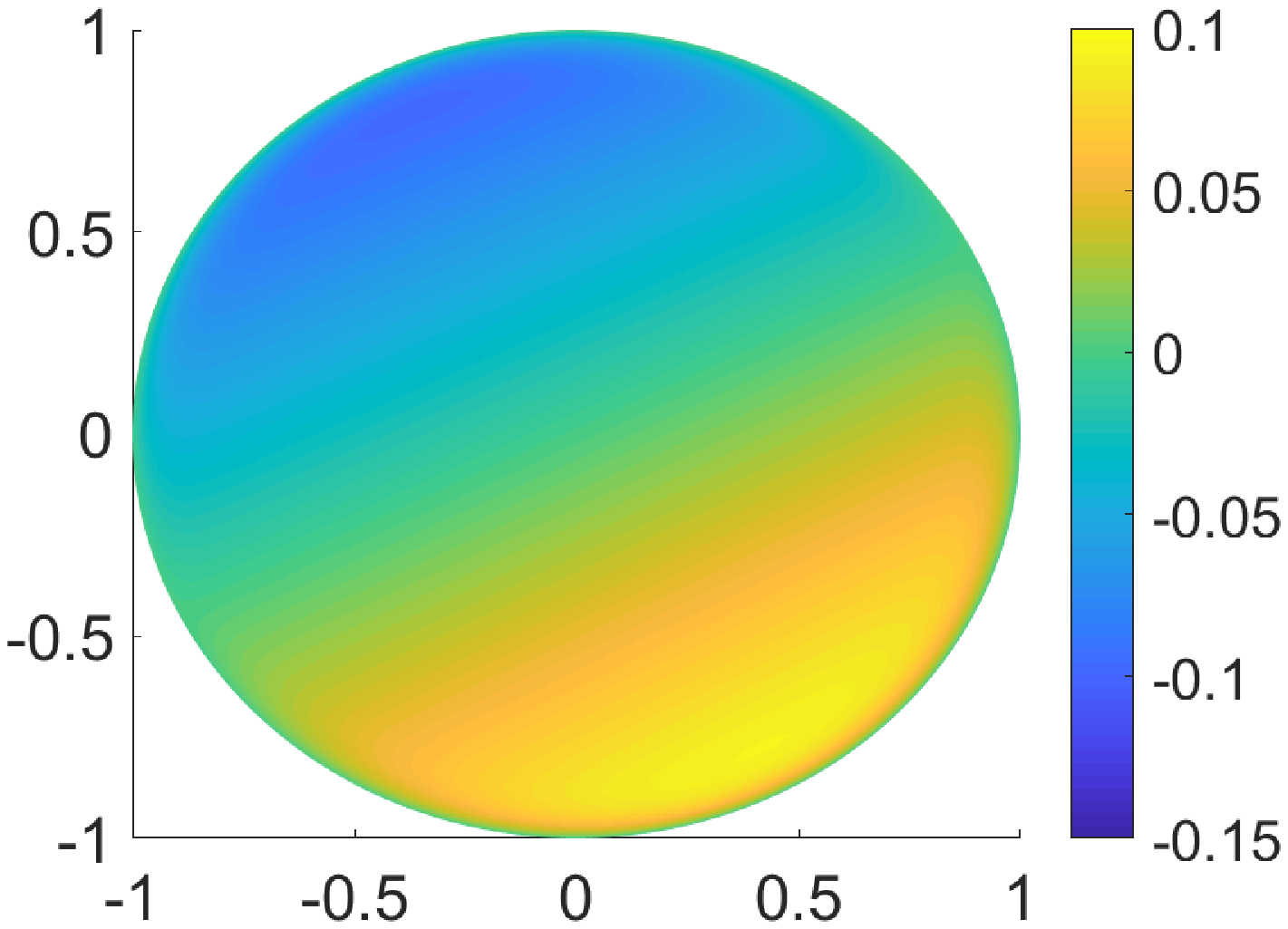}}
\subfigure[$\phi$ from EN model]{\includegraphics[width=3.1in]{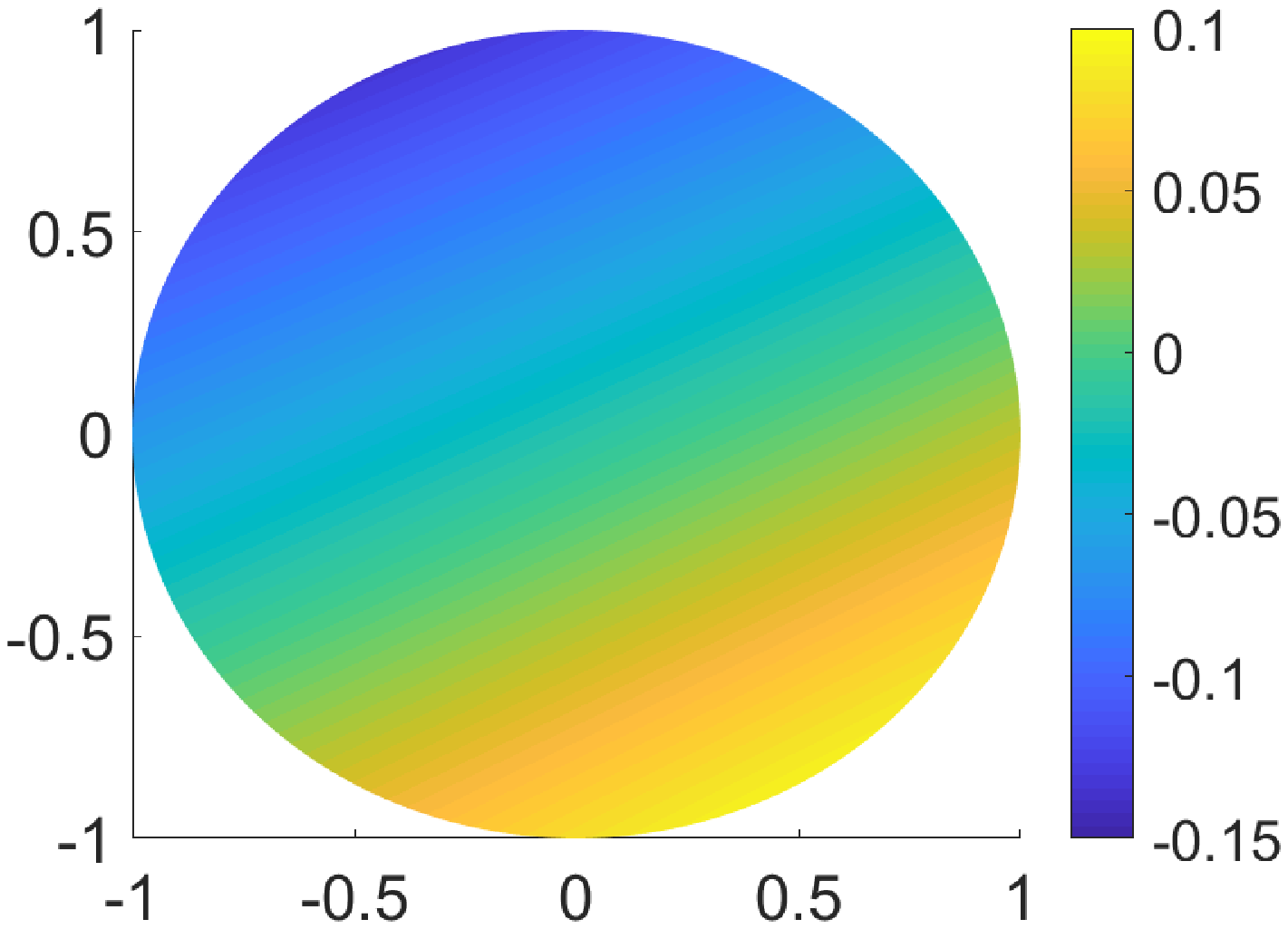}}
\caption{\label{figure5} Comparison of electric potentials $\psi(r,\theta,t)$ from PNP system and $\phi(r,\theta,t)$ from EN model with present condition (\ref{eq27_1}) at $t=0.5$. }
\end{center}
\end{figure}

In the simulation, finite element method with refined mesh (as in previous example) is used for original system  (\ref{eq5},\ref{eq6},\ref{Eq71}). Standard mesh is used for EN system (\ref{eq9},\ref{eq10}) together with effective boundary conditions in (\ref{eq40}). For boundary condition (\ref{eq40}), linearized implicit scheme is used to treat the $\partial_t $ term, while explicit scheme is used to treat $\partial_\theta$ term. Figure \ref{figure4} shows the comparison between $p(r,\theta,t)$ from PNP system and $c(r,\theta,t)$ from EN model, and Figure \ref{figure5} shows the comparison between $\psi(r,\theta,t)$ from PNP system and $\phi(r,\theta,t)$ from EN model at $t=0.5$, which is almost at steady state. They show that the approximate solutions $c(x,t)$ and $\phi(x,t)$ agree very well with exact solutions. The maximum errors of $c,\phi$ for some bulk region $r\in[0,0.5]$ at $t=0.5$ are respectively $4.2 \times 10^{-5} $ and $0.017$. Again, the EN system allows for relatively large mesh and time step sizes, and hence the computational time is greatly reduced, i.e., about 4.4 hours for the original PNP and 17 minutes for the EN system on the same computer (Processor: 4GHz, i76700K; Memory: 32GB).

 \setcounter{equation}{99} 

\section{EN model for action potential propagation}
\label{sec4}

As a concrete example, we consider the problem of propagation of an action potential along a neuronal axon.  This problem was first investigated in \cite{hodgkin1952} by a cable model. Later many works have simulated it in many cases \cite{cooley1966,pods2013}  and have attempted to recover the cable model based on PNP system and other assumptions \cite{qian1989,leonetti1998,mori2009}. We refer to the book \cite{malmivuo1995} for a good summary of cable model.  In this section, we first formulate the problem by using a PNP system and then derive an EN model. Then we present the simulations based on PNP system and the EN model to show the effectiveness of EN model.

\subsection{The formulation}
\label{sec4a}

\begin{figure}[h]
\begin{center}
\hspace{-1cm}\includegraphics[width=3.1in]{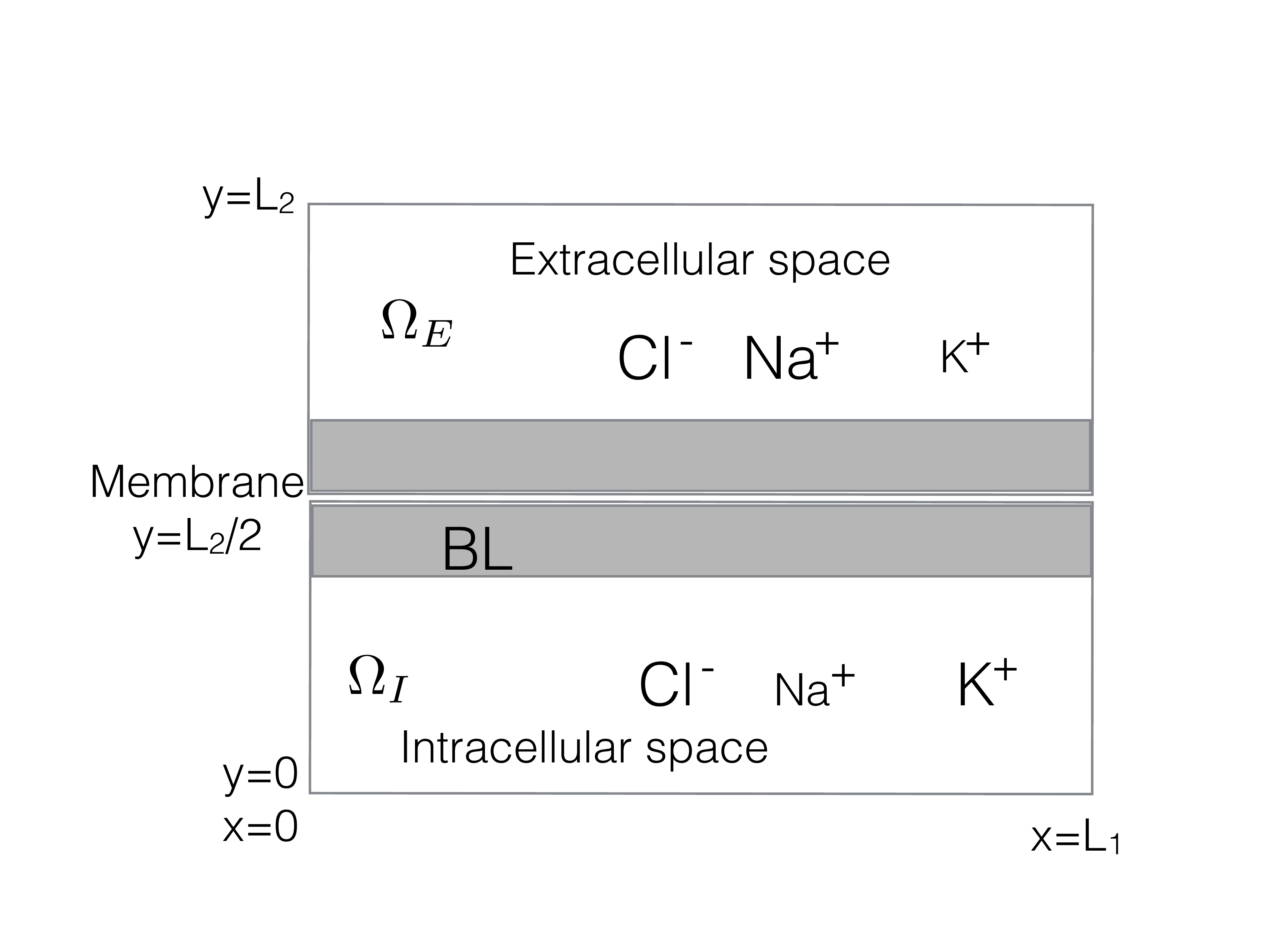}
\caption{\label{fig2} Sketch of the domain $\Omega$.}
\end{center}
\end{figure}

Here, we follow the formulation based on PNP system in \cite{pods2013}.  Due to symmetry of the axon, the problem is treated as a 2D problem. The domain $\Omega$ is a rectangular domain, with a membrane in the middle to separate the extracellular region and intracellular region. In Cartesian coordinates (see Figure \ref{fig2}), $\Omega$ is given by $(x,y) \in [0,L_1]\times [0,L_2]$, where $y$-direction is normal to the membrane. The membrane is the middle line $y=L_2/2$,  the lower region $\Omega_I =[0,L_1] \times [0,L_2/2)$ is the intracellular space and the upper region $\Omega_E = [0,L_1] \times (L_2/2,L_2]$ is the extracellular space. Only three basic ions (sometimes called bioions) $\mathrm{Na}^{+},\mathrm{K}^{+},\mathrm{Cl}^{-}$ are considered (fixed negative charge are incorporated into $\mathrm{Cl}^{-}$ ion as approximation), and LEN condition in bulk region is valid in this biological application.

We first formulate the original system in dimensional form. Let $p_i$ ($i=1,2,3$) denote ion concentrations of $\mathrm{Na}^{+},\mathrm{K}^{+},\mathrm{Cl}^{-}$, with valences $z_1=z_2 =1,z_3=-1$. The dimensional PNP system for $p_i$ and electric potential $\psi$ in domain $\Omega$ except the membrane is given by (\ref{eq1}), and we recall (see definition of parameters below (\ref{eq1}))
\begin{equation}
\label{eq100}
\begin{aligned}
&- \epsilon_0 \epsilon_r \Delta \psi  = e_0N_A\left(\sum_{k=1}^3 z_k p_k \right) ,\\
&\partial_t p_i = - \nabla \cdot \mathbf{J}_{p_i}= D_i \nabla \cdot  \left( \nabla p_i +\frac{e_0}{k_B T} z_i p_i  \nabla \psi\right),
\end{aligned}
\end{equation}
where $i=1,2,3$, and we have assumed the same relative permittivity $\epsilon_r$ for extracellular and intracellular regions.

The membrane at $y=L_2/2$ is described by Hodgkin-Huxley model \cite{hodgkin1990}, in order to simulate action potential for neuronal axon. Thus, the dimensional relation for the current through membrane/ion channel, from intracellular region to extracellular region, is
\begin{equation}
\label{eq101}
\begin{aligned}
& I_i =  G_{p_i} (x) (V_m - E_i),\quad i=1,2,3,
\end{aligned}
\end{equation}
or in terms of flux at $y=L_2/2$
\begin{equation}
\label{eq102}
\begin{aligned}
z_i e_0 N_A J_{p_i}^y \equiv&-z_i e_0 N_A D_i  \left(\partial_{y}p_i +\frac{e_0}{k_B T} z_i p_i \partial_{y}\psi\right) \\
=& G_{p_i} \left(\psi_I - \psi_E -  \frac{k_B T}{z_i e_0} \ln \frac{p_{iE}}{p_{iI}}\right),
\end{aligned}
\end{equation}
where $G_{p_i}$ is the conductance for ion $p_i$ and possibly depends on variable $x$ (e.g., myelinated axon), $E_i$ is the Nernst potential of ion $p_i$, $V_m = \psi_I-\psi_E$ is the membrane potential, and superscript $y$ in $J_{p_i}^y$ means the flux component in $y$-direction. Hereafter, subscripts $I$ and $E$ denote the values or limit values at the membrane $y = L_2/2$ from intracellular and extracellular regions respectively.  For the part of the axon without myelin sheath, the conductances $G_{p_i}$ depend on the membrane potential $V_m$. Following \cite{song2018,pods2013}, we set
\begin{equation}
\label{eq103}
\begin{aligned}
& G_{p_1} \equiv G_{\mathrm{Na}} = \bar{G}_{\mathrm{Na}} m^3 h +G_{\mathrm{Na},\mathrm{leak}}, \\
& G_{p_2} \equiv G_{\mathrm{K}} = \bar{G}_{\mathrm{K}} n^4 + G_{\mathrm{K},\mathrm{leak}},\\
& G_{p_3}\equiv G_{\mathrm{Cl}} =0,
\end{aligned}
\end{equation}
where $\bar{G}_{\mathrm{Na}},\bar{G}_{\mathrm{K}},G_{\mathrm{Na},\mathrm{leak}},G_{\mathrm{K},\mathrm{leak}}$ are some constant given in Appendix \ref{appendix_B}, and $n,m,h$ (associated with potassium channel activation, sodium channel activation, and sodium channel inactivation) depend on $V_m$ and are governed by a dynamic system in Appendix \ref{appendix_C}.

Suppose the membrane has a small thickness $h_m$ and relative permittivity $\epsilon_r^m$, and assume there are no ions in membrane. Thus, the electric potential is linear inside membrane. To complete the formulation, the other two jump conditions on the membrane $y=L_2/2$ are
\begin{equation}
\label{eq104}
\begin{aligned}
& \left.\epsilon_r \partial_{y}\psi \right|_{y=\frac{L_2}{2}\pm} = \epsilon_r^m \frac{\psi_E - \psi_I}{h_m},
\end{aligned}
\end{equation}
where $\frac{L_2}{2}\pm$ mean limits at membrane from upper and lower regions.

\subsection{Non-dimensionalization}
\label{sec4b}

In this subsection, we present the dimensionless PNP formulation combined with {the} HH model, and prescribe some suitable initial and boundary conditions.
We adopt the following scalings 
\begin{equation}
\label{eq105}
\begin{aligned}
& \tilde{\psi} = \frac{\psi}{k_B T/e_0}, \quad \tilde{p}_i = \frac{p_i}{p_0},\\
& \tilde{x} = \frac{x}{L_2}, \quad \tilde{y} = \frac{y}{L_2}, \quad \tilde{h}_m = \frac{h_m}{L_2}, \\
& \tilde{D}_i = \frac{D_i}{D_0}, \quad \tilde{t} = \frac{t}{L_2^2/D_0},\quad \tilde{G}_{p_i}= \frac{G_{p_i}}{G_0},
\end{aligned}
\end{equation}
where the length scale $L_2$ is adopted as in \cite{song2018} so that it gives the correct time scale for action potential, $p_0$ is the typical concentration of ions, $D_0$ is the typical diffusion constant, and typical conductance $G_0$ is defined by $G_0 = p_0 D_0 e^2 N_A/(k_B T L_2)$. All the parameter values and typical values are given in Appendix \ref{appendix_B}. In the following, we will remove the tilde, and still use the same notations but they represent dimensionless quantities.

The dimensionless PNP system in $\Omega = [0,L_1/L_2] \times [0,1]$ is given by (as in (\ref{eq2}))
\begin{equation}
\label{eq106}
\begin{aligned}
&- \epsilon^2 \Delta \psi  = \sum_{i=1}^n z_i p_i,\\
&\partial_t p_i = - \nabla \cdot \mathbf{J}_{p_i}= D_i \nabla \cdot ( \nabla p_i + z_i p_i  \nabla\psi),
\end{aligned}
\end{equation}
together with the conditions on interface $y=1/2$, 
\begin{equation}
\label{eq107}
\begin{aligned}
\left.z_i J_{p_i}^y \right|_{y=\frac12 \pm} \equiv  & -\left.z_i D_i  \left(\partial_{y}p_i + z_i p_i \partial_{y}\psi\right)  \right|_{y=\frac12 \pm}\\
=& G_{p_i} \left(\psi_I - \psi_E -  \frac{1}{z_i} \ln \frac{p_{iE}}{p_{iI}}\right),
\end{aligned}
\end{equation}
and
\begin{equation}
\label{eq108}
\begin{aligned}
\left.\epsilon^2 \partial_{y}\psi \right|_{y=\frac{1}{2}\pm} = C_m (\psi_E -\psi_I), ~~ C_m = \frac{\epsilon_m^2}{h_m},
\end{aligned}
\end{equation}
where $C_m$ is the dimensionless capacitance of membrane. In this system, the dimensionless parameters $\epsilon$ and $\epsilon_m$ are defined by
\begin{equation}
\label{eq109}
\begin{aligned}
& \epsilon = \sqrt{\frac{\epsilon_0 \epsilon_r k_B T}{e_0^2 N_A p_0 L_2^2}}, \quad  \epsilon_m = \sqrt{\frac{\epsilon_0 \epsilon_r^m k_B T}{e_0^2 N_A p_0 L_2^2}},
\end{aligned}
\end{equation}
which are given in Appendix \ref{appendix_B}.

We use typical bulk concentrations as the initial values (see Appendix \ref{appendix_B}) at $t=0$, then we have
\begin{equation}
\label{eq110}
\begin{aligned}
& p_1(x,y,0) = 1,\quad p_2(x,y,0) = 0.04,\\
& p_3 (x,y,0) = 1.04, ~~ \mathrm{in} ~~ \Omega_E = [0,\frac{L_1}{L_2}] \times (1/2,1],
\end{aligned}
\end{equation}
and
\begin{equation}
\label{eq111}
\begin{aligned}
& p_1(x,y,0) = 0.12,~~ p_2(x,y,0) = 1.25,\\
& p_3 (x,y,0) = 1.37, \quad \mathrm{in}\quad  \Omega_I=  [0,\frac{L_1}{L_2}] \times [0,1/2).
\end{aligned}
\end{equation}
For the boundary conditions, we adopt Dirichlet conditions on the top boundary (cf. Figure \ref{fig2})
\begin{equation}
\label{eq112}
\begin{aligned}
&\psi (x,1,t) = 0,\quad p_1(x,1,t) = 1,\\
& p_2(x,1,t) = 0.04,\quad p_3 (x,1,t) = 1.04,\\
\end{aligned}
\end{equation}
and zero-flux conditions on other boundaries
\begin{equation}
\label{eq113}
\begin{aligned}
&\frac{\partial \psi}{\partial y}(x,0,t) = 0,\quad  J_{p_i}^y(x,0,t) =0,\\
&\frac{\partial \psi}{\partial x}(0,y,t) = 0,\quad  J_{p_i}^x(0,y,t) =0,\\
&\frac{\partial \psi}{\partial x}(L_1/L_2,y,t) = 0,\quad  J_{p_i}^x(L_1/L_2,y,t) =0,\\
\end{aligned}
\end{equation}
where  $i=1,2,3$. The above system is coupled with the dynamic system for $m,h,n$ in Appendix \ref{appendix_C}, which determines the conductances $G_{p_i}$ in (\ref{eq107}) by (\ref{eq103}) on the membrane.

\subsection{The EN model with effective flux conditions}

By (\ref{eq50}), the EN equations for $c_1,c_2, \phi$ are given by
\begin{equation}
\label{eq114}
\begin{aligned}
&\partial_t c_i = - \nabla \cdot \mathbf{J}_{c_i}= D_i  \nabla \cdot ( \nabla c_i + z_i c_i  \nabla \phi),\\
&\sum_{i=1}^3 z_i D_i  \nabla \cdot ( \nabla c_i + z_i c_i  \nabla \phi) =0,
\end{aligned}
\end{equation}
where $i = 1,2$, $z_1=z_2=1,z_3=-1$ and $c_3 = c_1 + c_2$. The outer boundary $\partial \Omega$ lies in bulk region, so associated boundary conditions are easily derived from (\ref{eq112},\ref{eq113}), and we have
\begin{equation}
\label{eq115}
\begin{aligned}
&\phi (x,1,t) = 0,\, c_1(x,1,t) = 1,\, c_2(x,1,t) = 0.04,\\
& J_{c_k}^y(x,0,t) =0, \quad J_{c_k}^x(0,y,t) =0, \\
& J_{c_k}^x(L_1/L_2,y,t) =0, 
\end{aligned}
\end{equation}
where $k=1,2,3$.

As illustrated  in Figure \ref{fig2},  there are BLs at two sides of membrane. Then, we need to propose approximate jump conditions at middle interface for bulk quantities $c_{iI},\phi_I,c_{iE},\phi_E$ ($i=1,2$), where subscripts $I,E$ indicate the limit values at interface $y=1/2$ from intracellular (lower) and extracellular (upper) regions.  Based on previous results in Theorems 1 and 2, we first note that $\eta = x, \xi = \pm (y-1/2), g=1$ in the theorems and obtain the following 12 conditions
\begin{equation}
\label{eq116}
\begin{aligned}
&~~G_{p_i} \left(\psi_I - \psi_E -  \frac{1}{z_i} \ln \frac{p_{iE}}{p_{iI}}\right) \\
& = z_i \left( J_{c_i,E}^y + \epsilon \partial_t F_{iE} - \epsilon D_i \partial_x (F_{iE} \partial_x \mu_{iE})\right),\\
&~~G_{p_i} \left(\psi_I - \psi_E -  \frac{1}{z_i} \ln \frac{p_{iE}}{p_{iI}}\right) \\
& = z_i \left( J_{c_i,I}^y - \epsilon \partial_t F_{iI} + \epsilon D_i  \partial_x (F_{iI} \partial_x \mu_{iI})\right),\\
&\ln c_{iE} + z_i \phi_E + \frac{\epsilon J_{c_i,E}^y }{D_i} f_{iE}
= \ln p_{iE} + z_i \psi_E , \\
&\ln c_{iI} + z_i \phi_I - \frac{\epsilon J_{c_i,I}^y }{D_i} f_{iI} = \ln p_{iI} + z_i \psi_I,
\end{aligned}
\end{equation}
where $i=1,2,3$, $c_{3I} = c_{1I} + c_{2I},c_{3E} = c_{1E} + c_{2E}$ and we have defined
\begin{equation}
\label{eq117}
\begin{aligned}
& \mu_{is} = \ln c_{is} + z_i \phi_{s} \\
 & F_{is}= F_i(c_{1s},c_{2s},\phi_s-\psi_s),  \\
 & f_{is}= f_i(c_{1s},c_{2s},\phi_s-\psi_s), \quad s=I,E,
\end{aligned}
\end{equation}
where $F_i$ and $f_i$ are given by (\ref{eqa3},\ref{eqa6}). From Theorem 3, (\ref{eq108}) and (\ref{eqa8}), we get
\begin{equation}
\label{eq118}
\begin{aligned}
& C_m ({\psi_E - \psi_I}) =  \epsilon \sqrt{2 c_{3E}} \left(e^{(\phi_E - \psi_E)/2} -  e^{(\psi_E - \phi_E)/2}\right),\\
&C_m ({\psi_E - \psi_I})  = -\epsilon \sqrt{2 c_{3I}} \left(e^{(\phi_I - \psi_I)/2} -  e^{(\psi_I - \phi_I)/2}\right).
\end{aligned}
\end{equation}

From the definition (\ref{eq103}) and the data in Appendix \ref{appendix_B}, the conductances are small, i.e., $G_{p_i} \le O(\epsilon)$. Then, one can simplify the conditions in (\ref{eq116}) by neglecting higher order $O(\epsilon^2)$ terms, and we obtain the effective flux conditions at interface
\begin{equation}
\label{eq119}
\begin{aligned}
 & z_i J_{c_i,E}^y =  G_{p_i} \left(\phi_I - \phi_E -  \frac{1}{z_i} \ln \frac{c_{iE}}{c_{iI}}\right) \\
 &\quad - z_i \epsilon \partial_t F_{iE} + \epsilon D_i z_i \partial_x (F_{iE} \partial_x \mu_{iE}) ,\\
 & z_i J_{c_i,I}^y = G_{p_i} \left(\phi_I - \phi_E -  \frac{1}{z_i} \ln \frac{c_{iE}}{c_{iI}}\right) \\
&\quad + z_i \epsilon \partial_t F_{iI} - \epsilon D_i z_i \partial_x (F_{iI} \partial_x \mu_{iI}),\\
\end{aligned}
\end{equation}
The $\partial_t$ terms account for the ion accumulation in Boundary layer like a nonlinear capacitor \cite{song2018}, and the  $\partial_x$ terms account for the spacial variations along boundary. To summarize, the final EN model consists of (\ref{eq114},\ref{eq115}) and interface conditions (\ref{eq118},\ref{eq119}).

\noindent {\bf Remark 5.} By linearization according to small $\phi_I-\psi_I$ and $\phi_E-\psi_E$, we get from (\ref{eq117},\ref{eqa3},\ref{eq118}) that
\begin{equation}
\label{eq120}
\begin{aligned}
 &\epsilon z_i F_{iI} \approx C_m \lambda_{iI} V_m, \\
 & V_m = \psi_I - \psi_E \approx \phi_I - \phi_E,
 \end{aligned}
\end{equation}
where
\begin{equation}
\label{eq121}
\begin{aligned}
 \lambda_{iI} = \frac{c_{iI}}{\sum_{k=1}^3 c_{kI}}.\\
\end{aligned}
\end{equation}
Summation of fluxes in (\ref{eq119}) implies
\begin{equation}
\label{eq122}
\begin{aligned}
 & \sum_{i=1}^3 z_i J_{c_i,I}^y - \sum_{i=1}^3 G_{p_i} \left(\phi_I - \phi_E -  \frac{1}{z_i} \ln \frac{c_{iE}}{c_{iI}}\right)  \\
 &\approx C_m \partial_t V_m - C_m \sum_{i=1}^3 D_i \partial_x \left( \lambda_{iI} V_m \partial_x \mu_{iI} \right).
\end{aligned}
\end{equation}
Physically, the first term is the current from bulk region, the second term is Hodgkin-Huxley flux model (with bulk quantities), the right-hand side represents a capacitor and spacial variation along membrane.  One can further recover the classic cable model by adopting suitable scaling for variable $x$, which is left for future study.

\subsection{Numerical simulation}

In this subsection, we present numerical results using both the original PNP system and the present EN model. The computation is divided into two steps, first we generate a resting state, and second we simulate the propagation of action potential. We will study two case, i.e., axons with and without myelin sheath.

\begin{figure}[h]
\begin{center}
\subfigure[Dynamics of membrane potential $V_m$] {\includegraphics[width=2.6in]{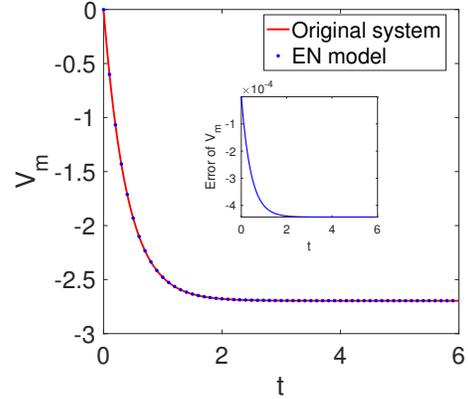}}
\subfigure[Distribution of electric potential at $t=6$]{\includegraphics[width=2.6in]{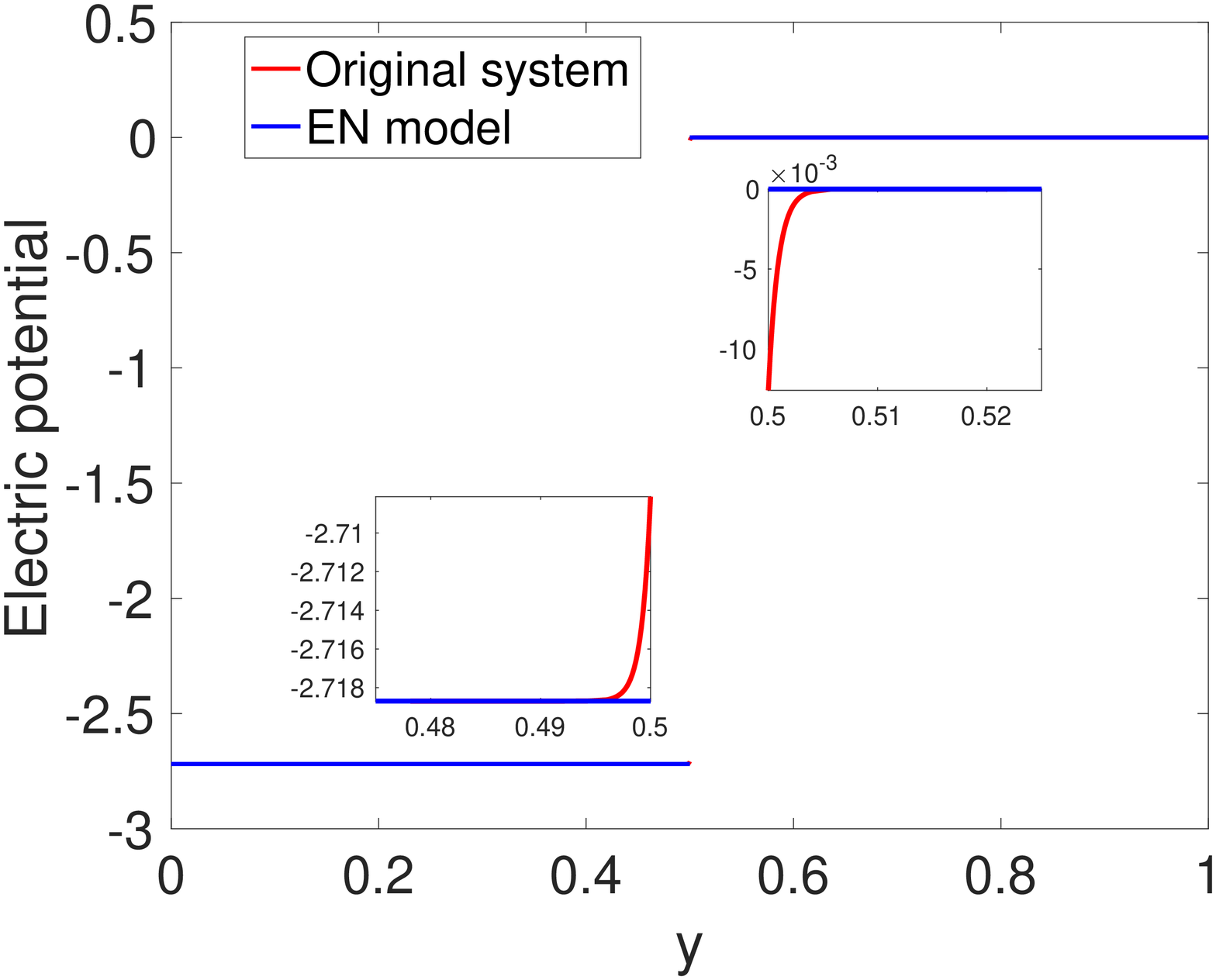}}
\caption{\label{figure7} Numerical results of original system (red) and electro-neutral (EN) model (blue) to generate the resting state in step 1.}
\end{center}
\end{figure}

First, we study unmyelinated axon. The length of axon is much larger than the typical scale of cell \cite{pods2013,ford2015}, and the domain is set to be $\Omega = [0,2000] \times [0,1]$. In step 1, to generate a resting state, we use the conductances in (\ref{eq103}) with equilibrium values for $n,m,h$ given in (\ref{eqc4}). In the computation, we use a 1D code for $y$-direction, since the problem is uniform in $x$. For the original model, finite element method with non-uniform fixed mesh is adopted, where mesh size varies from $1.6\times10^{-4}$ near the BL to $3.3\times10^{-2}$ in the bulk. Uniform mesh with mesh size $3.3\times10^{-2}$ is adopted in EN model.  Flux of sodium ion $J_{p_1}^y$ is negative, i.e., from $\Omega_E$ to $\Omega_I$, while flux of potassium ion $J_{p_2}^y$ is positive. After certain period, e.g., at $t=6$, the net flux across membrane tends to 0, i.e., $\left.J_{p_1}^y+ J_{p_2}^y\right|_{y=1/2} =0$, which is set as the resting state. Figure \ref{figure7}(a) shows the dynamics of membrane potential $V_m=\psi_I-\psi_E$ for both the original model and the new EN model, and the two solutions agree very well with each other (error is shown in the figure). Figure \ref{figure7}(b) shows the distributions of electric potential $\psi$ for the original system and $\phi$ for the EN model, at resting state $t=6$. They agree very well in the domain except the BL. The resting potential is calculated as 
\begin{equation}
\label{eq123}
\begin{aligned}
& \left.V_m\right|_{t=6} =\left.\psi_I-\psi_E \right|_{t=6} \approx -2.7,\\
& V_r = \frac{k_B T}{e_0} \left.(\psi_I-\psi_E)\right|_{t=6}  \approx  -65 \, \mathrm{mV}.
\end{aligned}
\end{equation}

\begin{figure}[h]
\begin{center}
\subfigure[$\Delta t = 10^{-4}$] {\includegraphics[width=2.5in]{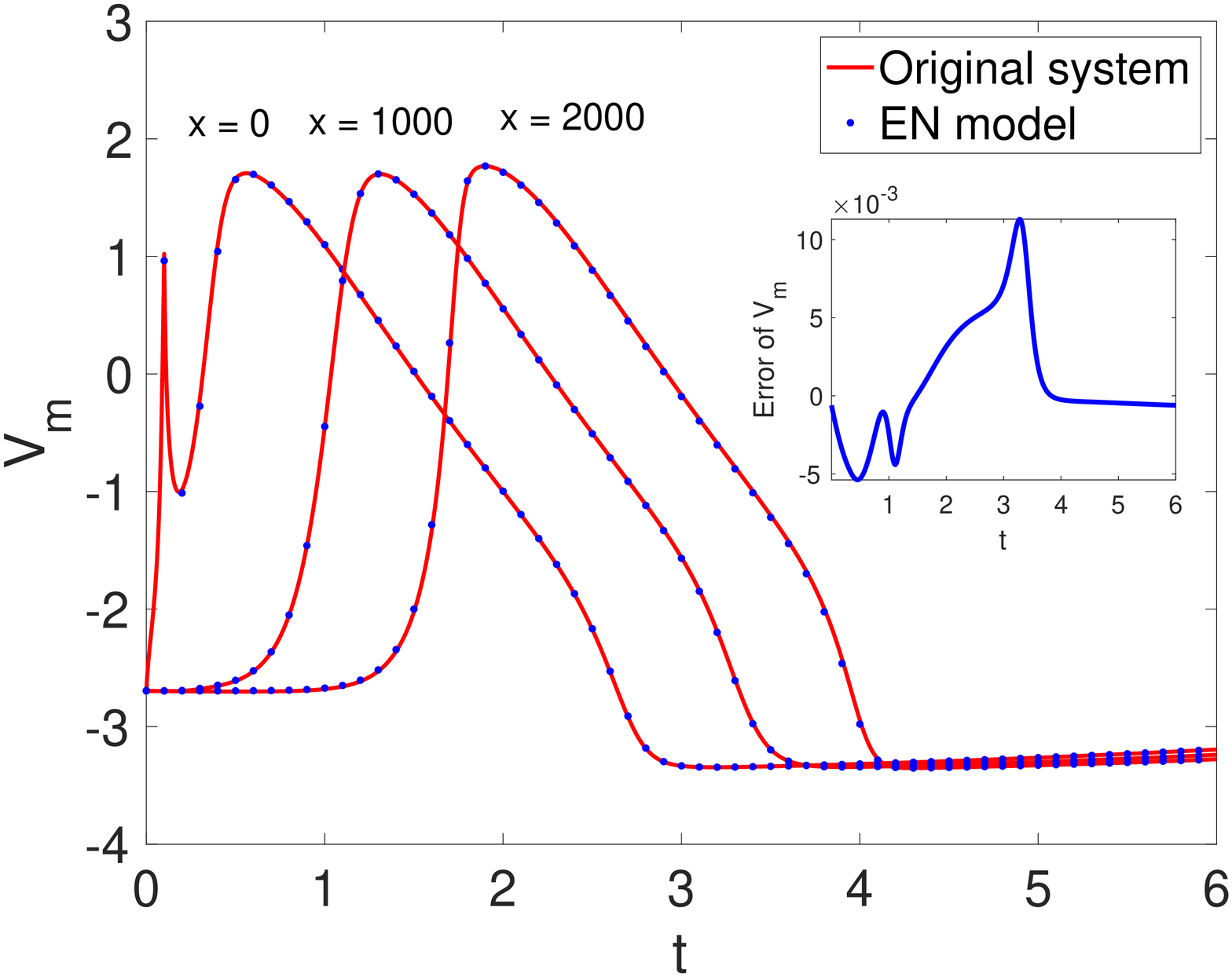}}
\subfigure[$\Delta t = 5\times10^{-4}$] {\includegraphics[width=2.5in]{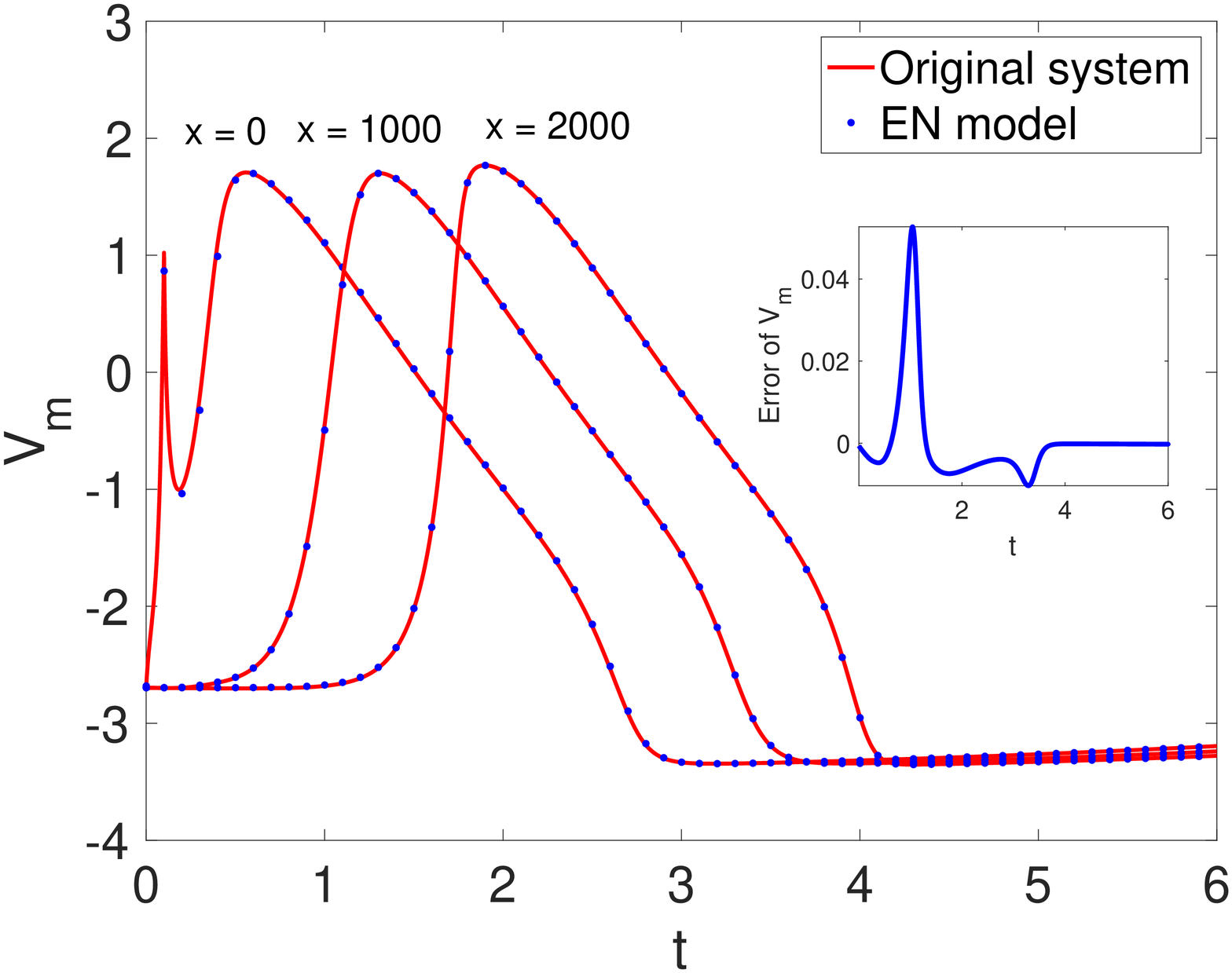}}
\subfigure[$\Delta t = 10^{-3}$] {\includegraphics[width=2.5in]{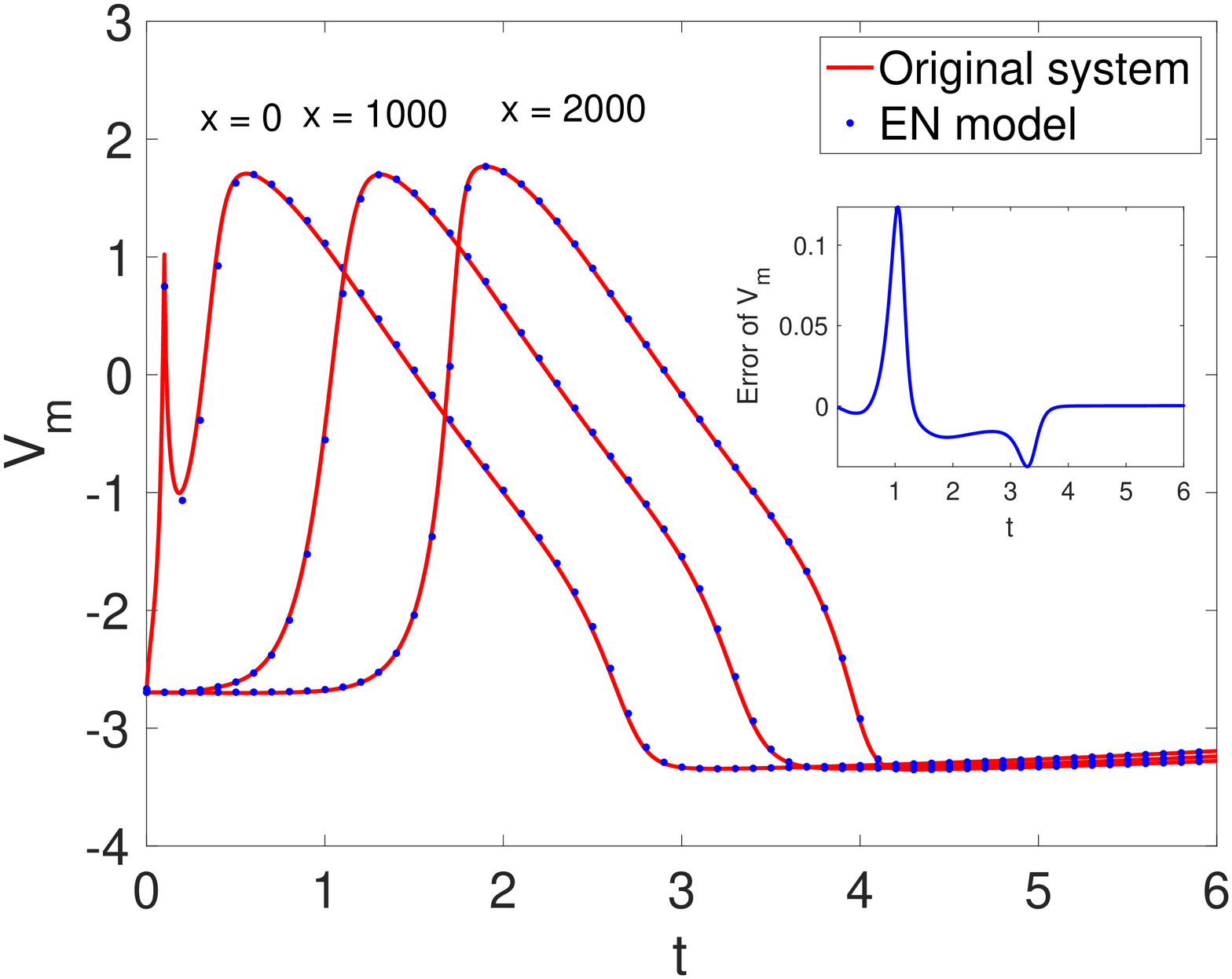}}
\caption{\label{figure8} Numerical results of original system with $\Delta t = 10^{-4}$ and EN model with three different time step sizes, and the error of $V_m$ for $x=1000$ is shown in the figures.}
\end{center}
\end{figure}

In step 2, to simulate the propagation of action potential \cite{pods2013}, we use the conductances in (\ref{eq103}), where $n,m,h$ depend on membrane potential $V_m$ and their dynamics are given in Appendix \ref{appendix_C}.  To initiate the action potential near $x=0$ on the membrane, we increase the conductance of $G_{p_1}(x)$ by modifying $\bar{G}_{Na}$ (to the value $0.6$) in the interval $x\in [0,60]$ for the time period $0<t<0.1$. This allows extra influx of sodium ion into $\Omega_I$ and hence generates the action potential.  In the computation, finite element method is used for both original system and EN model. For original system, implicit scheme for nonlinear terms is adopted to avoid some stability issues due to small parameter $\epsilon$, and the ``exact" numerical solution is calculated with time step size $\Delta t = 10^{-4}$. For EN model, there is no BL and it allows for relatively larger time step sizes. We try three implementations for EN model with different time step sizes $\Delta t = 10^{-4},5\times10^{-3},10^{-3}$. Figure {\ref{figure8}} shows the dynamics of membrane potential $V_m(x,t) = \psi_I - \psi_E$ at different locations of membrane obtained by using the original model and the new EN model. Action potential first occurs at $x=0$, and then propagates to the positive $x$. The error of $V_m$ at $x=1000$ is also shown in the figure, indicating good agreement of the two models. The computation time and the maximum error for $V_m$ are listed in Table {\ref{table5}} compared with the exact results for original system. It indicates that it costs 56 hours for original system, while the computation time is greatly reduced with EN model, where all computations are done on the same computer (Processor: 4GHz, i76700K; Memory: 32GB). So the EN model is more efficient with acceptable accuracy. The conductance velocity is defined as the velocity that action potential (the electric signal) travels along the axon. In this example, it is estimated as 1.3 m/s in dimensional quantities, which is the same order as usual estimates \cite{pods2013}. This is slightly larger than that in \cite{pods2013}, since the length of axon is not long enough and the boundary effect at $x=0,2000$ influences the velocity.

\begin{figure}[h]
\begin{center}
\subfigure[3/4 myelinated] {\includegraphics[width=2.5in]{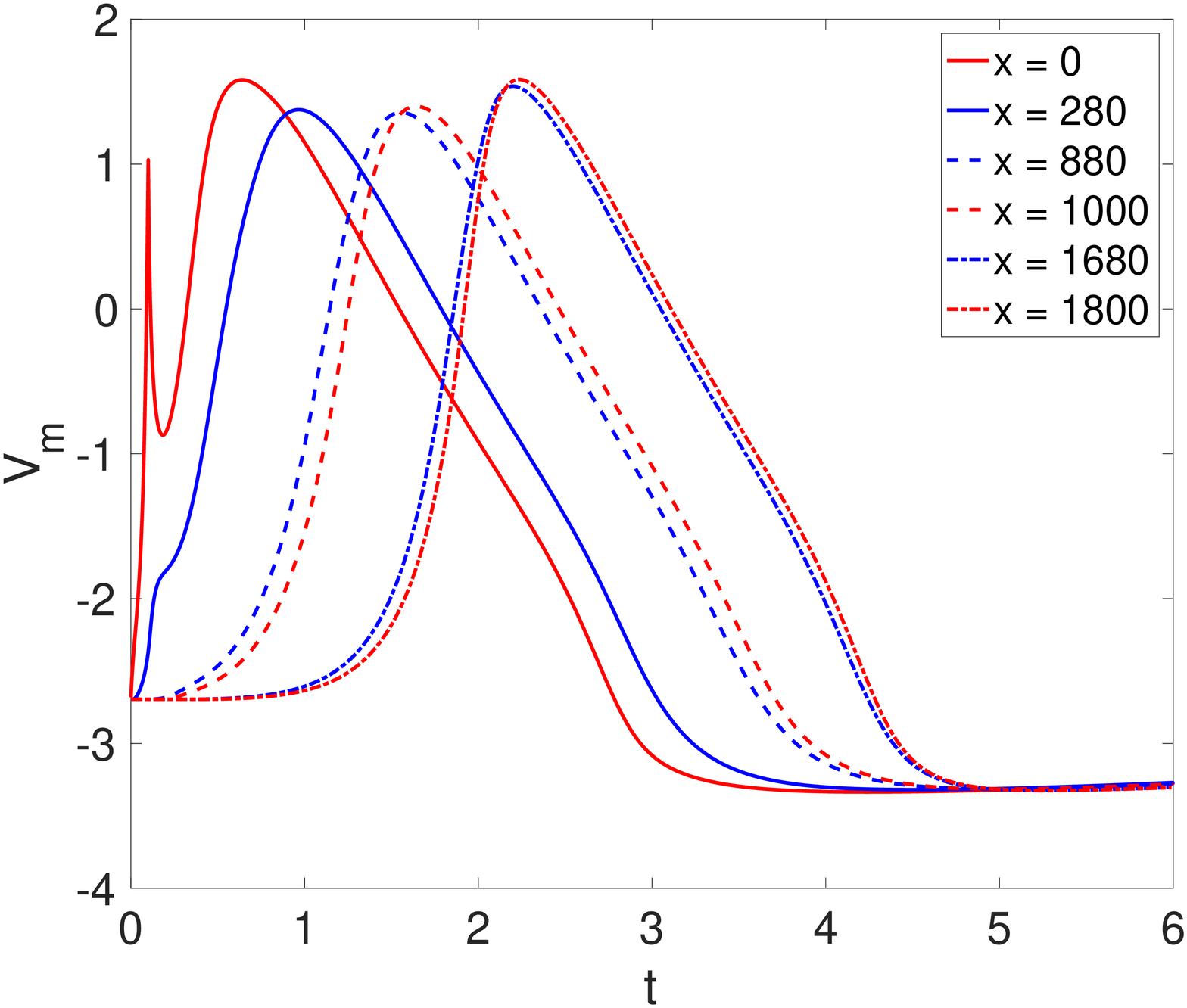}}
\subfigure[9/10 myelinated] {\includegraphics[width=2.5in]{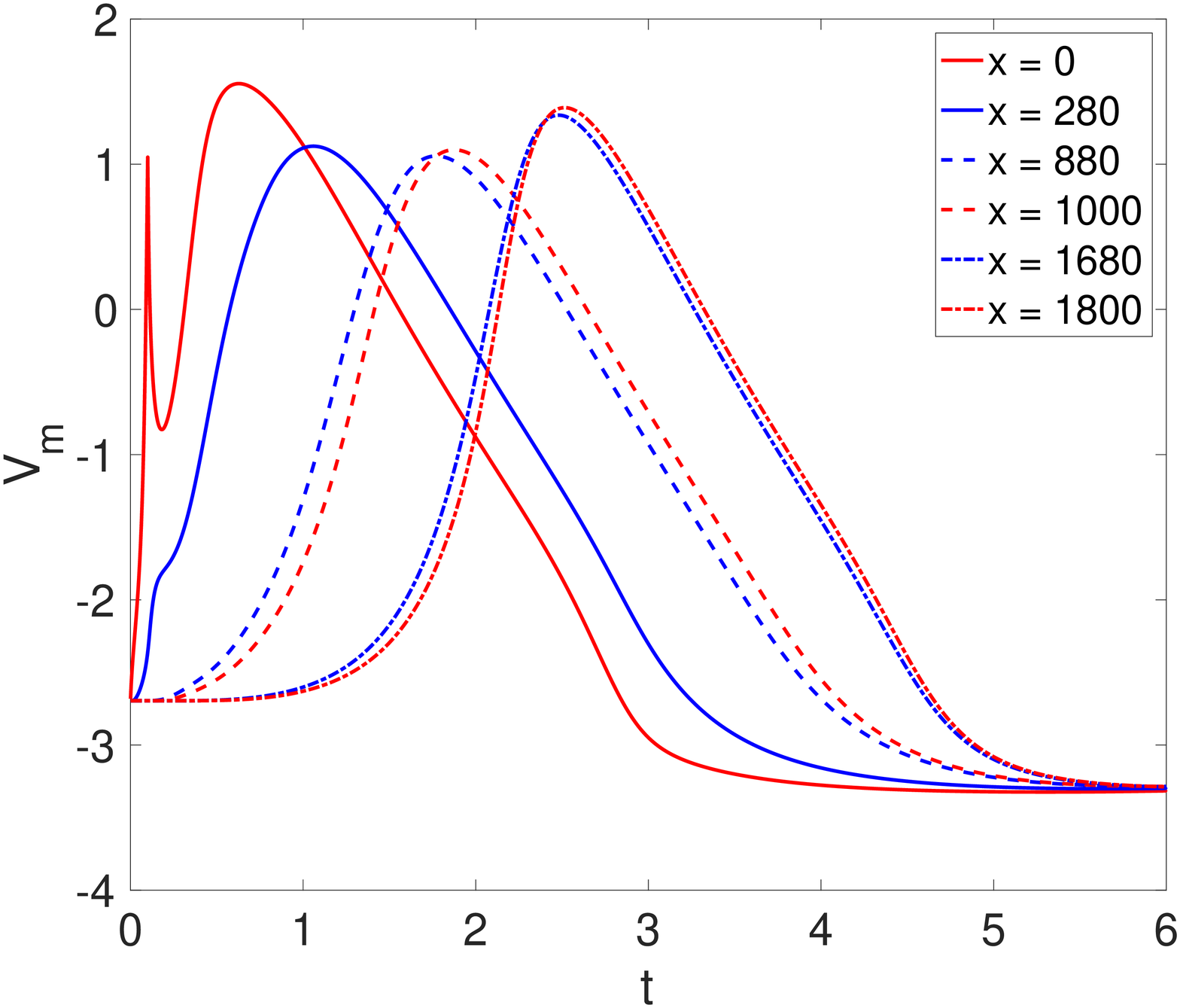}}
\caption{\label{figure9} Numerical results of propagation of action potential for myelinated axon.}
\end{center}
\end{figure}

\begin{table*}
\begin{center}
\begin{tabular}{ c|c|c|c|c}
    \hline\hline
 & Original system $\Delta t=10^{-4}$  & EN model, $\Delta t=10^{-4}$  &  EN model, $\Delta t=5\times10^{-4}$  & EN model, $\Delta t=10^{-3}$   \\ \hline
Error & $ - $   &  $0.01$  & $0.05$ &  $0.12$\\
Time & 56 hours      & 20 hours      & 3 hours  & 1.9 hours \\
\hline
\end{tabular}
\end{center}
\caption {\label{table5}Comparison of computation time between original system and EN model, and the maximum error for membrane potential $V_m$ in EN model.}
\end{table*}

In the second case, we consider the myelinated axon, where conductances $G_{p_i}(x)$ are nonzero at only unmyelinated parts (typically the nodes). By \cite{ford2015}, each segment between nodes is roughly $100-300$ (scaled by $1\mu m$), here it is set to be $200$. To see the qualitative effect, we increase the portion of myelinated part in each segment of axon, where the portions 3/4 and 9/10 are tested. Figure \ref{figure9} shows the propagation of action potential for myelinated axon, calculated with EN model. In the figures, blue and red curves represent action potential  at some locations for myelinated and unmyelinated parts respectively. The action potential is initiated at the $x=0$, weakens at myelinated parts, and reinforces a little at unmyelinated part (node) of each segment. For the 3/4 myelinated axon, the peak values of action potential gradually decrease from about 1.6 at $x=0$ to about 1.4 at $x=1000$ and recovers to about 1.6 at $x=1800$. For the 9/10 myelinated axon, the peak values of action potential decrease from about 1.6 at $x=0$ to about 1.1 at $x=1000$ and then increase to about 1.4 at $x=1800$. In the latter case, the tested axon is not long enough for the signal (action potential) to fully recover to its original strength.

\section{Conclusions}

In this work, we have investigated a 2D dynamic PNP system with various boundary conditions, and have derived the corresponding EN system with effective boundary conditions. In the case of Dirichlet boundary conditions, the effective conditions can be considered as generalization of continuity of electrochemical potential. For flux conditions, we derived a physically correct effective conditions by keeping some essential high-order terms, which are important in many biological applications. The effective conditions for the general multi-ion species case involves elliptic integrals, and these extra terms of elliptic integrals account for the accumulation of ions in the BL and the spacial variation along boundary. We have validated our EN models with several examples and demonstrated the effectiveness of the EN system with the implementation of the well-known Hodgkin-Huxley model for propagation of action potential on axon. 

As a next step, for the biological example in Section \ref{sec4} we will analyse the reduction from EN system to the classic cable model, under some consistent assumptions, and then we will see the effects of different levels of approximations. We also plan to extend our approach to modified PNP system where size effect of the ions are included.

\appendix

\section{Expressions of functions in Theorems 1,2,3}
\label{appendix_A}

For some special cases, the explicit expressions for $F_i,f_i$ and relation (\ref{eq74}) are available. For the previous case $z_1=1,z_2=-1$, we recover the result
\begin{equation}
\label{eqa1}
\begin{aligned}
& F_1(c_{10},\phi_0-\psi_0) =  \sqrt{2 c_{10}} (e^{(\phi_0 - \psi_0)/2} -1),\\
& F_2(c_{10},\phi_0-\psi_0) =  \sqrt{2 c_{10}} (e^{(\psi_0 - \phi_0)/2} -1).
\end{aligned}
\end{equation}
For the case $z_1=2,z_2 =-1$, we get
\begin{equation}
\label{eqa2}
\begin{aligned}
F_1 (c_{10},\phi_0-\psi_0)=&  \sqrt{\frac{c_{10}}{2}} \left[ e^{\frac{\phi_0 - \psi_0}{2}} \sqrt{e^{(\phi_0 - \psi_0)} +2} -\sqrt{3} \right],\\
F_2 (c_{10},\phi_0-\psi_0)= & \sqrt{2 c_{10}} \left( \sqrt{1+ 2 e^{(\psi_0 - \phi_0)}} -\sqrt{3} \right).
\end{aligned}
\end{equation}
For the 3-ion case with $z_1=1,z_2=1,z_3=-1$, we have
\begin{equation}
\label{eqa3}
\begin{aligned}
F_j (c_{10},c_{20},\phi_0-\psi_0)= & \sqrt{\frac{c_{j0}}{c_{10}+c_{20}}}  \sqrt{2 c_{j0}} \left(e^{\frac{\phi_0 - \psi_0}{2}} -1\right),\\
F_3 (c_{10},c_{20},\phi_0-\psi_0) = & \sqrt{2 (c_{10}+c_{20})} (e^{(\psi_0 - \phi_0)/2} -1),
\end{aligned}
\end{equation}
where $j=1,2$.

For the case $z_1= 1,z_2 =-1$, we have 
\begin{equation}
\label{eqa4}
\begin{aligned}
f_1 (c_{10},\phi_0-\psi_0)= & \frac{\sqrt{2} (e^{(\psi_0 - \phi_0)/2} -1)}{  c_{10}^{3/2}}  , \\
f_2 (c_{10},\phi_0-\psi_0)= &\frac{\sqrt{2} (e^{(\phi_0 - \psi_0)/2} -1)}{c_{10}^{3/2}}.
\end{aligned}
\end{equation}
For the case $z_1=2,z_2=-1$, we get
\begin{equation}
\label{eqa5}
\begin{aligned}
f_1 = &  \frac{\sqrt{2+ e^{\phi_0 - \psi_0}} (1+ 2 e^{\phi_0 - \psi_0} )e^{\frac{3}{2} (\psi_0 - \phi_0)}  - 3\sqrt{3}}{3 \sqrt{2} c_{10}^{3/2}},\\
f_2 = &  \frac{\mathrm{arcsinh} \left(e^{(\phi_0 - \psi_0)/2}/\sqrt{2} \right)  - \mathrm{arccsch}(\sqrt{2})}{ \sqrt{2} c_{10}^{3/2}}.
\end{aligned}
\end{equation}
For the case with $z_1=1,z_2=1,z_3=-1$, we have
\begin{equation}
\label{eqa6}
\begin{aligned}
f_j (c_{10},c_{20},\phi_0-\psi_0)= & \frac{\sqrt{2} (e^{(\psi_0 - \phi_0)/2} -1)}{  c_{j0} \sqrt{c_{10}+c_{20}}}  , \\
f_3 (c_{10},c_{20},\phi_0-\psi_0)= &\frac{\sqrt{2} (e^{(\phi_0 - \psi_0)/2} -1)}{  (c_{10}+c_{20})^{3/2}} ,
\end{aligned}
\end{equation}
where $j=1,2$.

For the case $z_1= 1,z_2 =-1$, the relation (\ref{eq74}) becomes
\begin{equation}
\label{eqa7}
\begin{aligned}
\psi_0 -\tilde{\psi}_0= \frac{\gamma}{\epsilon} \sqrt{2 c_{10}} \left( e^{(\phi_0 - \psi_0)/2} - e^{(\psi_0 - \phi_0)/2}  \right).
\end{aligned}
\end{equation}
For the case $z_1= 1,z_2 =1,z_3 = -1$, it becomes
\begin{equation}
\label{eqa8}
\begin{aligned}
\psi_0 -\tilde{\psi}_0= \frac{\gamma}{\epsilon} \sqrt{2 c_{30}} \left( e^{(\phi_0 - \psi_0)/2} - e^{(\psi_0 - \phi_0)/2}  \right),
\end{aligned}
\end{equation}
where $c_{30} = c_{10} + c_{20}$ by EN condition.

\section{The data used in Section \ref{sec4}}
\label{appendix_B}

The data are mainly from papers \cite{hodgkin1990,pods2013} and the book \cite{malmivuo1995}. The temperature in \cite{hodgkin1990} is set to be $6.3^o C$, so  we get $T = 279.45 \, \mathrm{K}$. The other constants are
\begin{equation}
\label{eqb1}
\begin{aligned}
& k_B  = 1.38 \times 10^{-23} \,\mathrm{J}/\mathrm{K}, \quad N_A = 6.022 \times 10^{23} /\mathrm{mol}, \\
& e_0= 1.602 \times 10^{-19} \,\mathrm{C},~ \epsilon_0 = 8.854 \times 10^{-12} \, \mathrm{C}/(\mathrm{V\cdot m}).
\end{aligned}
\end{equation}
The typical bulk concentrations for $\mathrm{Na}^+,\mathrm{K}^+, \mathrm{Cl}^-$ are
\begin{center}
\begin{tabular}{cccc}
 & $p_1,\mathrm{Na}^+$  & \quad $p_2,\mathrm{K}^+$ & \quad $p_3,\mathrm{Cl}^-$\\
Extracellular & $100\, \mathrm{mM}$ & $4\, \mathrm{mM}$ &$104\, \mathrm{mM}$\\
Intracellular & $12\, \mathrm{mM}$ & $125\, \mathrm{mM}$  &  $137\, \mathrm{mM}$\\
\end{tabular}
\end{center}
\noindent which are used as initial conditions (scaled by $p_0$ below). Some typical values are (diffusivity of $\mathrm{Cl}^-$ is from \cite{bob2014})
\begin{equation}
\label{eqb2}
\begin{aligned}
& \epsilon_r  =80, \quad \epsilon_r^m= 2,\quad h_m = 5 \mathrm{nm}, \\
& L_1 = 100\mu \textrm{m}\sim 10 \textrm{mm},\quad  L_2 = 1 \mu \mathrm{m},\\
& p_0 = 100\, \mathrm{mM}= 100 \, \mathrm{mol}/\mathrm{m}^3, \\
& D_0 = 10^{-5} \, \mathrm{cm}^2/\mathrm{s}= 10^{-9} \, \mathrm{m}^2/\mathrm{s}, \\
&D_1 = 1.33 D_0,~ D_2 = 1.96 D_0,~ D_3 = 2.03 D_0.
\end{aligned}
\end{equation}
The conductances are given by
\begin{equation}
\label{eqb3}
\begin{aligned}
& \bar{G}_{\mathrm{Na}} = 120\, \mathrm{mS}/\mathrm{cm}^2 = 1200 \, \mathrm{C}/(\mathrm{V\cdot s \cdot m^2}), \\
& \bar{G}_{\mathrm{K}} = 360 \, \mathrm{C}/(\mathrm{V\cdot s \cdot m^2}),\\
& \bar{G}_{\mathrm{Na},\mathrm{leak}} = 1.04 \, \mathrm{C}/(\mathrm{V\cdot s \cdot m^2}),\\
& \bar{G}_{\mathrm{K},\mathrm{leak}} = 4 \, \mathrm{C}/(\mathrm{V\cdot s \cdot m^2}).
\end{aligned}
\end{equation}
where leak conductances are set to ensure that resting potential is roughly 65 {mV}.

From the above data, we get
\begin{equation}
\label{eqb4}
\begin{aligned}
&\frac{k_B T}{e_0} \approx 24  \, \mathrm{mV} ,\quad \frac{L_2^2}{D_0} = 1\, \mathrm{ms},\\
& G_0= \frac{p_0 D_0 e^2 N_A}{k_B T L} \approx 400758 \, \mathrm{C}/(\mathrm{V\cdot s \cdot m^2}).
\end{aligned}
\end{equation}
For the dimensionless system we have 
\begin{equation}
\label{eqb5}
\begin{aligned}
&\epsilon= 1.33\times 10^{-3},\quad \epsilon_m =2.1\times 10^{-4}, \\
& h_m = 5 \times 10^{-3},\\
& D_1 = 1.33,\quad D_2 = 1.96, \quad D_3= 2.03,\\
& \bar{G}_{\mathrm{Na}} = 3  \times 10^{-3},\quad  \bar{G}_{\mathrm{K}} =9  \times 10^{-4},\\
& \bar{G}_{\mathrm{Na},\mathrm{leak}} = 2.6 \times 10^{-6},\quad \bar{G}_{\mathrm{K},\mathrm{leak}} = 1 \times 10^{-5}.
\end{aligned}
\end{equation}

\section{The dynamic system for $m,h,n$ in conductances}
\label{appendix_C}

The dynamics for $m,h,n$ in (\ref{eq103}) are given by \cite{malmivuo1995}
\begin{equation}
\label{eqc1}
\begin{aligned}
& \frac{dn}{dt} = \alpha_n (1-n) -\beta_n n,\\
& \frac{dm}{dt} = \alpha_m (1-m) -\beta_m m,\\
& \frac{dh}{dt} = \alpha_h (1-h) -\beta_h h.\\
\end{aligned}
\end{equation}
The coefficients depend on $V_m$ and are given by
\begin{equation}
\label{eqc2}
\begin{aligned}
& \alpha_n =\frac{1}{100} \frac{10-\bar{V}}{\left(e^{(10-\bar{V})/10} -1\right)},\quad 
\beta_n = \frac{1}{8 e^{\bar{V}/80}},\\
& \alpha_m = \frac{1}{10} \frac{25-\bar{V}}{\left(e^{(25-\bar{V})/10} -1\right)}, \quad \beta_m = 4 e^{-\bar{V}/18},\\
& \alpha_h= \frac{7}{100} e^{-\bar{V} /20}, \quad \beta_h = \frac{1}{e^{(30-\bar{V})/10}+1},
\end{aligned}
\end{equation}
where $\bar{V} = V_m - V_r$ and $V_r$ is some fixed resting potential. In above coefficients, the unit for $\bar{V}$ is millivolt. Theoretically, there is no singularity in above coefficients, but for computation when $\bar{V}$ is near $10$ or $25$, it is sensitive as denominator approaches 0. We can use the Taylor expansions in a small neighbourhood say $\delta=0.01$,
\begin{equation}
\label{eqc3}
\begin{aligned}
& \alpha_n(\bar{V} ) = \frac{1}{10} + \frac{\bar{V} -10}{200}+ \frac{(\bar{V} -10)^2}{12000},\quad |\bar{V}-10|<\delta,\\
& \alpha_m(\bar{V} ) =1+ \frac{\bar{V} -25}{20} + \frac{(\bar{V} -25)^2}{1200},\quad |\bar{V}-25|<\delta,
\end{aligned}
\end{equation}
and the error by choosing $\delta=0.01$ is at least at the order of $10^{-12}$.
With $\bar{V}=0$, we obtain the steady state solution
\begin{equation}
\label{eqc4}
\begin{aligned}
& n_\infty = \frac{4}{5e -1} \approx 0.3177,\\
& m_\infty = \frac{5}{8 e^{5/2} -3} \approx 0.05293,\\
& h_\infty= \frac{7 (1+e^3)}{107 + 7 e^3} \approx 0.5961,
\end{aligned}
\end{equation}
which are used to generate resting state and used as initial values of the time-dependent problem to simulate action potential.

For the dimensionless system in Section \ref{sec4b},  we still use the system (\ref{eqc1}) and will not scale the quantities in the coefficients (\ref{eqc2}), where the quantity $\bar{V}$ (in millivolts) is related to normalized membrane potential $V_m=\psi_I-\psi_E$ through
\begin{equation}
\label{eqc5}
\begin{aligned}
& \bar{V} =  \frac{k_B T}{e_0}(\psi_I-\psi_E) - V_r,
\end{aligned}
\end{equation}
and $V_r= -65$ mV is the resting potential in millivolts (see (\ref{eq123})).


\end{document}